\documentclass[journal]{IEEEtran}

\usepackage[numbers,sort&compress]{natbib}
\usepackage{url}
\usepackage{bm}
\usepackage{amsfonts}
\usepackage{amsmath}
\usepackage{amssymb}
\usepackage{amsthm}
\usepackage{color}
\usepackage{enumerate}
\usepackage{makecell}

\usepackage{array,color}
\usepackage{tabularx}
\usepackage{multirow}
\usepackage{booktabs}
\usepackage{makecell}  

\usepackage{graphicx}  
\usepackage{subfig}
\usepackage{epstopdf}
\usepackage{graphics}
\usepackage{epsfig}
\usepackage{psfrag}
\usepackage{overpic}

\usepackage{multicol}

 \usepackage{stfloats}
 \usepackage{float}   

 \usepackage{indentfirst} 
\usepackage{gensymb}

\usepackage{color, colortbl}

\begin{document}

\title{Preliminary Exploration on Digital Twin for Power Systems: Challenges, Framework, and Applications}

\author{Xing~He,~\IEEEmembership{Member,~IEEE},  Qian~Ai,~\IEEEmembership{Senior Member,~IEEE}, Robert~C. Qiu,~\IEEEmembership{Fellow,~IEEE}, Dongxia~Zhang
}
\maketitle
\begin{abstract}
Digital twin (DT) is one of the most promising enabling technologies for realizing smart grids.
Characterized by seamless and active---\textit{data-driven}, \textit{real-time}, and \textit{closed-loop}---integration between digital and physical spaces, a DT is much more than a blueprint, simulation tool, or cyber-physical system (CPS).
Numerous state-of-the-art technologies such as internet of things (IoT), 5G, big data, and artificial intelligence (AI) serve as a basis for DT. DT for power systems aims at \textit{situation awareness} and \textit{virtual test} to assist the decision-making on power grid operation and management under normal or urgent conditions.
This paper, from both science paradigms and engineering practice, outlines the backgrounds, challenges, framework, tools,  and possible directions of DT as a preliminary exploration.
To our best knowledge, it is also the \textit{first exploration} on DT in the context of power systems.
Starting from the fundamental and most frequently used power flow (PF) analysis, some typical application scenarios are presented.
Our work is expected to contribute some novel discoveries, as well as some high-dimensional analytics, to the engineering community. Besides, the connection of DT with big data analytics and AI may has deep impact on data science.
\end{abstract}

\begin{IEEEkeywords}
Digital twin, data-driven, real-time, closed-loop,  situation awareness, big data analytics, modeling
\end{IEEEkeywords}

\IEEEpeerreviewmaketitle
\section{Introduction}
\label{Introd}
\IEEEPARstart{M}{odern} power grid is one of the most complex engineering systems in existence; the North American power grid is recognized as the supreme engineering achievement in the 20th century \cite{doe2003grid}. The complexity of grids is ever increasing: 1) the evolution of grid networks, especially the expansion in size;  2) the penetration of renewable/distributed resources, flexible/controllable electronic components, or even prosumers with dual load-generator behavior~\cite{grijalva2011prosumer}; 3) the revolution of operation mechanisms, e.g., demand-side management; and 4) the  mechatronic disciplines (mechanics, electric and electronics) are realized in a more integrated way, and their interfaces will be more intertwined.
All these driving forces lead to a non-linear, diversified, hierarchical, and distributed power grid, which is hard to model and to analyse.

Digital twin (DT) is one of the most promising enabling technologies for realizing smart grids, especially for the construction of Ubiquitous SG-eIoT (Electric Internet of Things proposed by State Grid Corporation of China \cite{zhang2019UIoT}).
A DT refers to the digital representation of a real-world entity or system.  These DTs are linked to their real-world counterpart and are used to understand the state of the thing or system, respond to changes,  improve operation and add value.

DT is characterized by \textit{data-driven mode}, \textit{real-time interaction}, and \textit{closed-loop feedback}.
Automation and numerous state-of-the-art technologies such as internet of things (IoT), 5G,  drones, robots, edge computing, big data, and artificial intelligence (AI) serve as a basis for DT.
Organizations often get a quick start of DT, i.e., build simple DTs and put them into operation at the very beginning, and then evolve them over time in an \textit{active} and \textit{self-adaptive way}---evolve DTs with the accumulation of \textit{operation data}, \textit{practice feedbacks}, and \textit{subjective experience}.
It is quite different from the procedure that we build a physical model, in which a global designing, a deliberate start, and some assumptions and simplifications, are required in advance. \textit{This quick start makes DT much more accessible than conventional simulations}, e.g., Matpower, to engineering in practice.

It has been almost 16 years since the concept of DT was initially proposed in 2003 \cite{grieves2014digital}. To date, many DT applications have been successfully implemented in different industries and DT becomes an emerging market.
Tao et al. suggested 14 potential DT applications, such as product design, assembly, in a workshop \cite{Tao2018Digital}.
Gartner identified DT as one of the Top 10 Strategic Technology Trends of 2018 \cite{garfinkel2018gartner}.
The DT market is estimated to grow from USD 3.8 billion in 2019 to USD 35.8 billion by 2025, at a CAGR of 37.8\%. Major factors surging the demand for DT include increasing adoption of emerging technologies such as IoT and cloud \cite{markets2019DTmarktet}.

This paper, from both science paradigms and engineering practice, outlines the backgrounds, challenges, framework, tools,  and possible directions of DT for power systems (PSDT). To our best knowledge, it is also the \textit{first exploration} on DT in the context of power systems. Our preliminary exploration is expected to contribute some novel discoveries,  as well as some high-dimensional analytics, to the engineering community. Besides, the connection of DT with big data analytics and AI may has deep impact on data science.

\section{Background and Framework of PSDT}
\label{BakFram}
\subsection{From Twins to Digital Twins}
The concept of ``twins'' dates back to NASA's Apollo Program. In the program, at least two identical space vehicles were built, allowing engineers to \textit{mirror} conditions of the space vehicle during missions (analogous to Task 1, real-time situation awareness), and the vehicle remaining on earth is called the twin.
The twin was also used extensively for training during flight preparations (analogous to Task 2, ultra-time virtual test).
During a flight mission it was used to simulate alternatives on Earth-based model, where available flight data were used to mirror flight conditions as well as possible, and thus assist astronauts in orbit in critical situations.

Another well known example of a ``hardward'' twin is the Iron Bird, a ground-based engineering tool used in aircraft industries to incorporate, optimize and validate vital aircraft systems \cite{shafto2010draft}.
Due to the development of simulation technologies, the hardware parts in the Iron Bird are replaced by virtual models. This allows system designers to use the concept of an Iron Bird in earlier development cycles, even when some physical components are not yet available. Extending this idea further along all phases of the life cycle leads to a complete digital model of the physical system, the Digital Twin (DT).

\begin{figure}[htbp]
\centerline{
\includegraphics[width=.44\textwidth]{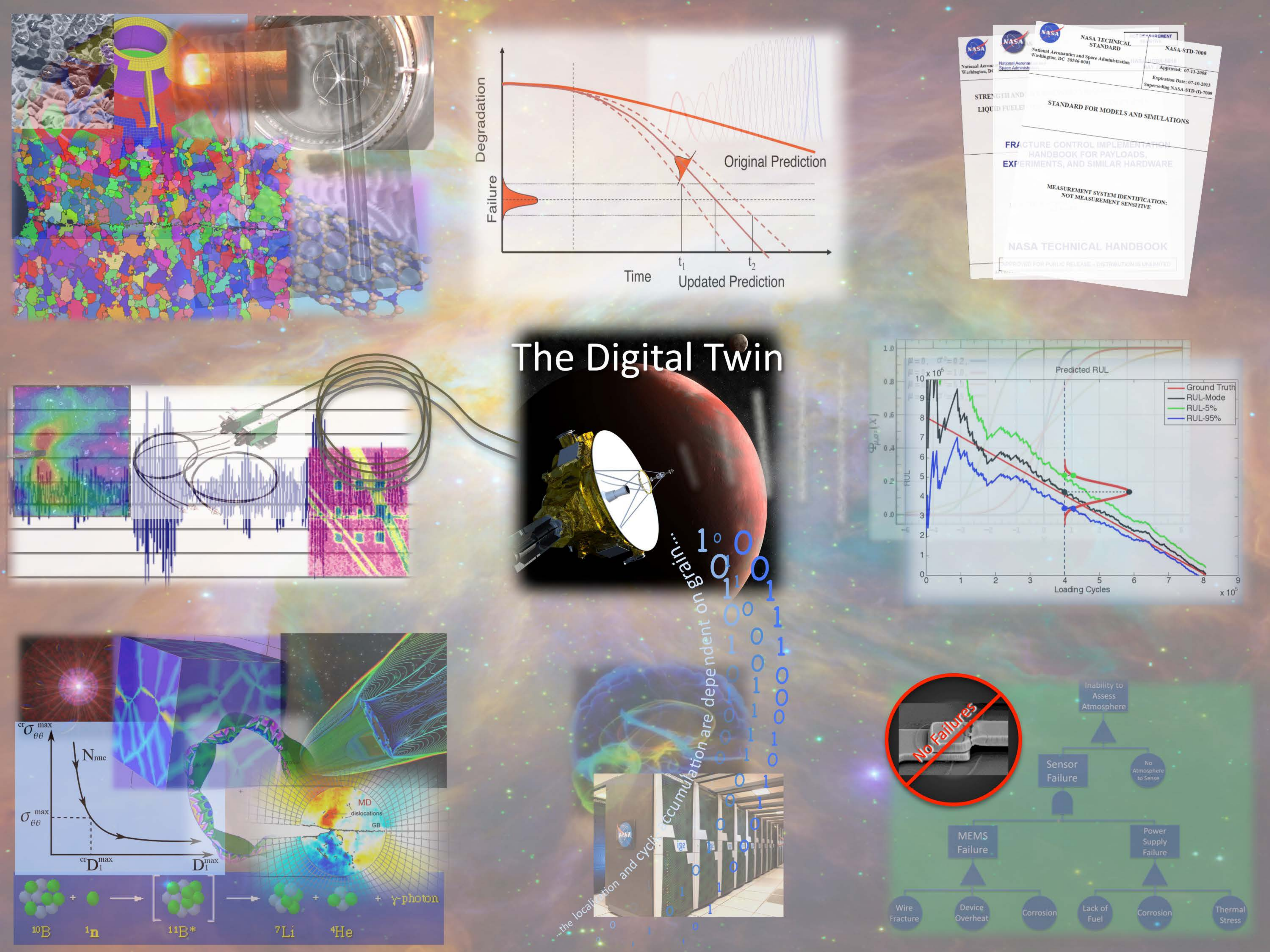}
}
\caption{Graphical Representation of DT Paradigm \cite{Glaessgen2012The}}
\label{fig: DTPara}
\end{figure}

The term DT was brought to the public for the first time in NASA's integrated technology roadmap \cite{shafto2012dt}:
\small{} \textit{A Digital Twin is an integrated multiphysics, multiscale simulation of a vehicle or system that uses the best available physical models, sensor updates, fleet history, etc., to mirror the life of its corresponding flying twin. The Digital Twin is ultra-realistic and may consider one or more important and interdependent vehicle systems, including propulsion/energy storage, avionics, life support, vehicle structure, thermal management/TPS, etc. Manufacturing anomalies that may affect the vehicle may also be explicitly considered.}\normalsize{}
Reference \cite{Glaessgen2012The} gives a graphical representation of some attributes of a DT as Fig. \ref{fig: DTPara}, and each of the nine subfigures  presents a narrative.

Our work explores PSDT. Rather than life cycle phases or manufacturing anomalies, PSDT is more concerned with two major tasks in the field of power systems: 1) \textit{real-time situation awareness} (SA), and 2) \textit{ultra-time virtual test}.

\subsection{Conventional SA and its Limitation}

SA is defined, according to \cite{endsley2011designing}, as  the \emph{perception} of the elements in an environment, the \emph{comprehension} of their meaning, and the \emph{projection} of their status in the near future.
Timely and accurate SA is essential for power system security. Inadequate SA is identified as one of the root causes  for the largest blackout in history---the 14 August 2003 Blackout in the United States and Canada \cite{us2004final}.
For a modern grid as stated in Sec. \ref{Introd}, SA is in urgent need of a new prominence.

\begin{figure}[htbp]
\centerline{
\includegraphics[width=.44\textwidth]{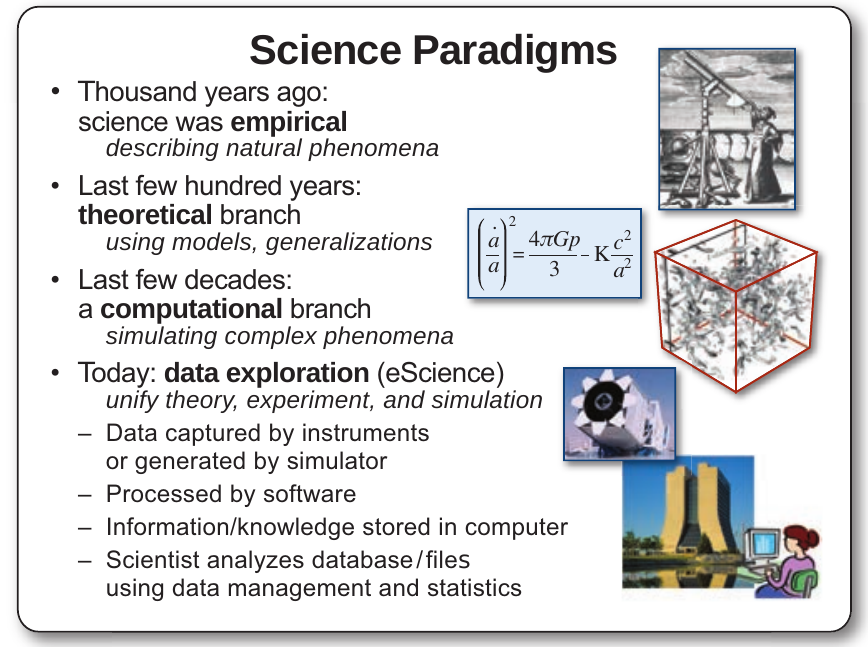}
}
\caption{Science Paradigms \cite{hey2009fourth}}
\label{fig:fourthp}
\end{figure}

Conventional model-based SA for power systems needs to be revisited. We would like to refer to Fig. \ref{fig:fourthp} in book \cite{hey2009fourth} as an illustration. Under the 2nd and 3rd paradigms (from last few hundred years to last few decades), our insight into the world is mainly based on physical models. For a power grid, we use equations, formulas, or simulations to describe the operation regulations and interaction mechanisms of each units.
This model-based mode cannot make full use of massive data due to its own limitations---it aims at a deterministic solution which is always in low-dimensional space (low dimension is not well compatible with high-dimensional data). Moreover, the assumptions and simplifications of system units (often small in size but large in number \cite{he2018SA}), and the increasing penetration of distributed energy resources (often susceptible to climate changing and usage lifetime \cite{he2019invisible}), will inevitably cause error; the error accumulations can hardly be addressed (described or analysed) with physical models or in low-dimensional space.

\subsection{Science Paradigms and Data-driven SA in High Dimension}
As shown in Fig.  \ref{fig:fourthp}, we are now entering the age of  the 4th-paradigm---data-intensive scientific discovery.
In this age, \textit{data-driven} is an alternative paradigm, and even becomes a trend as data become more and more accessible.
For smart grids, data-driven approaches become natural and stressing topics~\cite{bda2016tsg}.
Data-driven approaches are also characterized by model-free---we no longer heavily rely on physical models, and can handle the scenarios where the system topologies and network parameters are \textit{unreliable or even totally unavailable}.
Moreover, \textit{comparing data-driven results to model-based ones, we can obtain some insights for further analysis}.
This phenomenon has some connection with Task 2, virtual test, and will be demonstrated by the case studies in Sec. \ref{sec:vtjac}.

High in dimensionality, rather than large in number, is the keystone and difficulty of  data-driven SA designing.
High dimension means that the datasets are represented in terms of large random matrices. These data matrices can be viewed as data points in high-dimensional vector space of mathematics---each vector is very long.
Traditional data transformations, however, are often in the form of low dimension, such as one-dimensional Fourier Transformation (time domain to frequency domain), and three-dimensional Park Transformation (ABC to dq0).
In low-dimensional space, only two typical data matrices in the form of $\mathbf{X}\!\in\! {\mathbb{R}}^{N\times T}$ are at our disposal:  1)  $N,T$ are small, and 2) $N$ is small, $T$ is very large (compared with $N$).
Low dimension is not well compatible with high-dimensional data---low-dimensional tools are inadequate to a complex problem such as optimized operation or integrated planning for a modern grid, in which the behaviors and discipline of system units are closely intertwined.

\subsection{Big Data Analytics and AI for eScience Data}
\label{sec:BDAandAI}
The limitation of model-based and low-dimensional statistical algorithms to eScience data has greatly spurred the development of big data analytics and artificial intelligence (AI).
Modern grid operation is always accompanied with a temporal-spatial dataset---the dataset is in high-dimensional space, and in the form of time series. Temporal variations ($T$ sampling instants) are simultaneously observed together with spatial variations ($N$ signals).
\textit{Only in high-dimensional space do those statistical properties and benefits come out}, and it will be revealed by a virtual test in our case studies.

The extraction of statistical information  from this high-dimensional space is a challenge that does not meet the prerequisites of most mathematical tools.
Big data analytics and AI are employed to handle this challenge, and accordingly we select two tools: 1) random matrix theory (RMT), which is good at big data analytics, and 2) deep learning, which does well in massive data modeling.
Both of the tools have already made huge impacts on many engineering fields.
Compared to its model-based counterpart, DT is more compatible with or even more naturally connected to these tools.
PSDT realizes SA mainly \textit{by means of mining information from temporal-spatial dataset}.
\textit{This attribute of DT is marked as data-driven}.

\subsection{Data-driven Tools: RMT and Deep Learning}
\label{sec:RMTDL}
RMT has an advantage of transparency---unifying time and space through their ratio $c \!=\! T /N$, RMT deals with temporal-spatial data mathematically rigorously.
The goal of RMT is to understand the \textit{joint eigenvalue distribution} as the statistic analytics from big data in the asymptotic regime. In particular, high-dimensional analysis and visualization are treated as the functionals of the eigenvalue distributions.
For instance, Linear eigenvalue statistics (LESs) \cite{shcherbina2011central}, built from data matrices, follow Gaussian distributions for very general conditions, and other statistical variables are studied due to the latest breakthroughs in probability on the central limit theorems of those LESs \cite{qiu2015smart}.
The statistical properties of these variables are mostly derivable and provable. Moreover, RMT performs well with moderate-size data.

Deep learning  is the state-of-the-art algorithm in data science. Deep learning does learn some non-handcrafted features, so called deep features, from the massive labeled dataset without much prior knowledge, so that it can be generalized to different cases without making significant modifications. Moreover, the performance of the deep network model on fitting task and generalization task could be quantitative evaluated with test set, so as to ensure desired effects.


\subsection{Virtual Test and Framework of DT}
Besides data-driven SA, DT also allows for simulations of new ideas that can be tested virtually to determine environmental impact before implementation in the real world.
Software for automation can also be tested in advance using the virtual representation of the real system (i.e. ultra-time virtual test).

\begin{figure}[htbp]
\centerline{
\includegraphics[width=.44\textwidth]{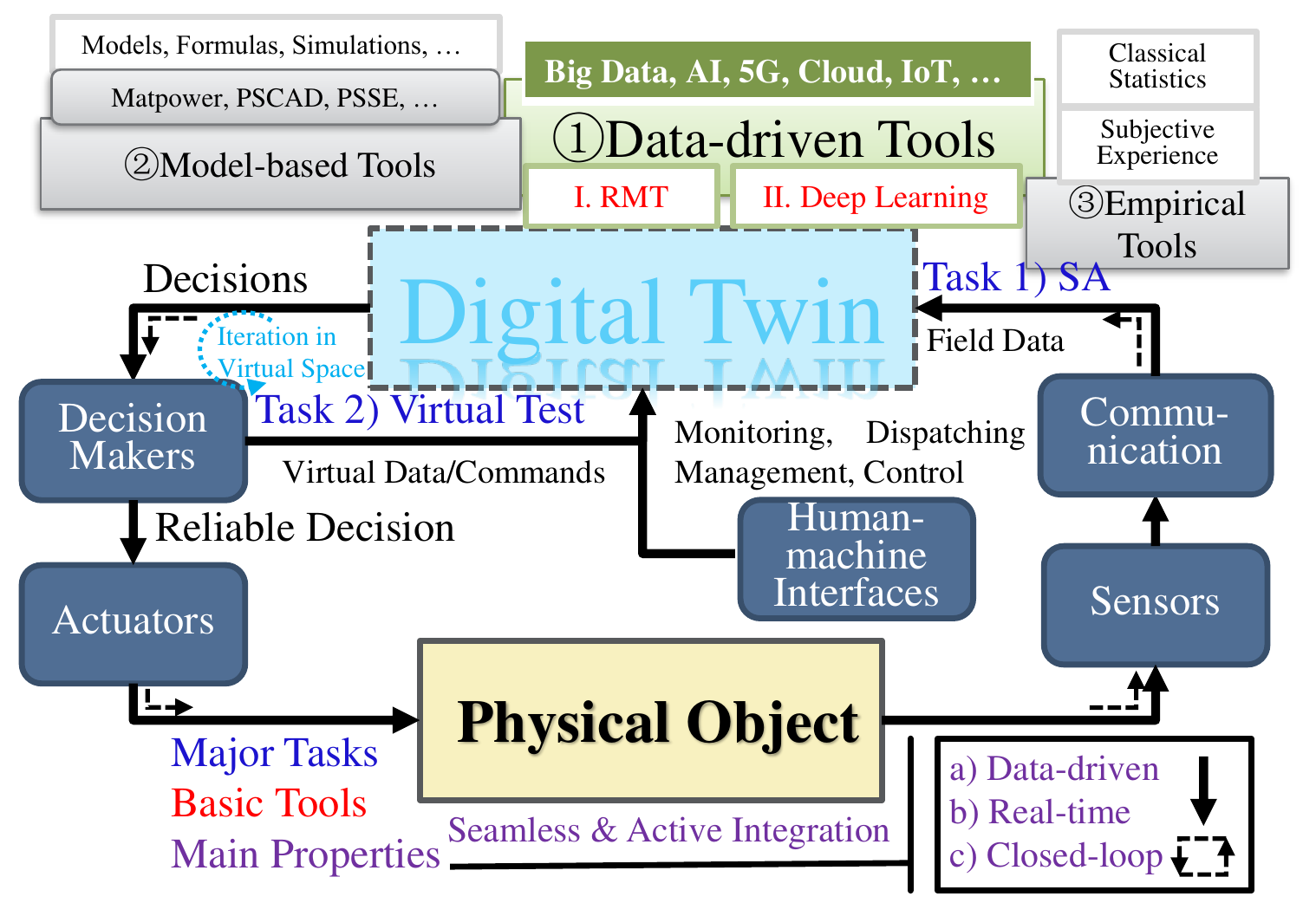}
}
\caption{Framework for DT of Power Systems}
\label{fig:DTFrame}
\end{figure}

Then we build the framework of PSDT, as shown in Fig. \ref{fig:DTFrame}.
The relationship among the PSDT, the physical grids, and the operators
are similar to the relationship, in the aforementioned Apollo Program, among the vehicle remaining on earth, the space vehicle during missions, and the astronauts in orbit.
PSDT helps the operators to understand the physical grid and to make a reliable decision in critical situations.

The arrow lines represent the data/information flow, which connects the physical object and its DT. Our two major tasks are deployed along these arrow lines: DT uses sampling spatial-temporal data from communication for SA, and uses virtual data/commands from decision makers for virtual test. These lines also form a closed loop, which means feedbacks are available for the tasks.
\textit{Characterized by real-time data/information flow and closed-loop feedbacks, the effectiveness of our tasks can be guaranteed.}
The involvement of human-machine interfaces provides an access to some artificial inputs, e.g., hyper parameters setting for a virtual test; it makes DT more flexible and intelligent in the virtual space.

\section{Power Flow Operation Formulation and Conventional Analysis}
\label{BakFram}
\subsection{Background of Power Grid Operation}
\begin{figure}[htpb]
\centering
\includegraphics[width=0.40\textwidth]{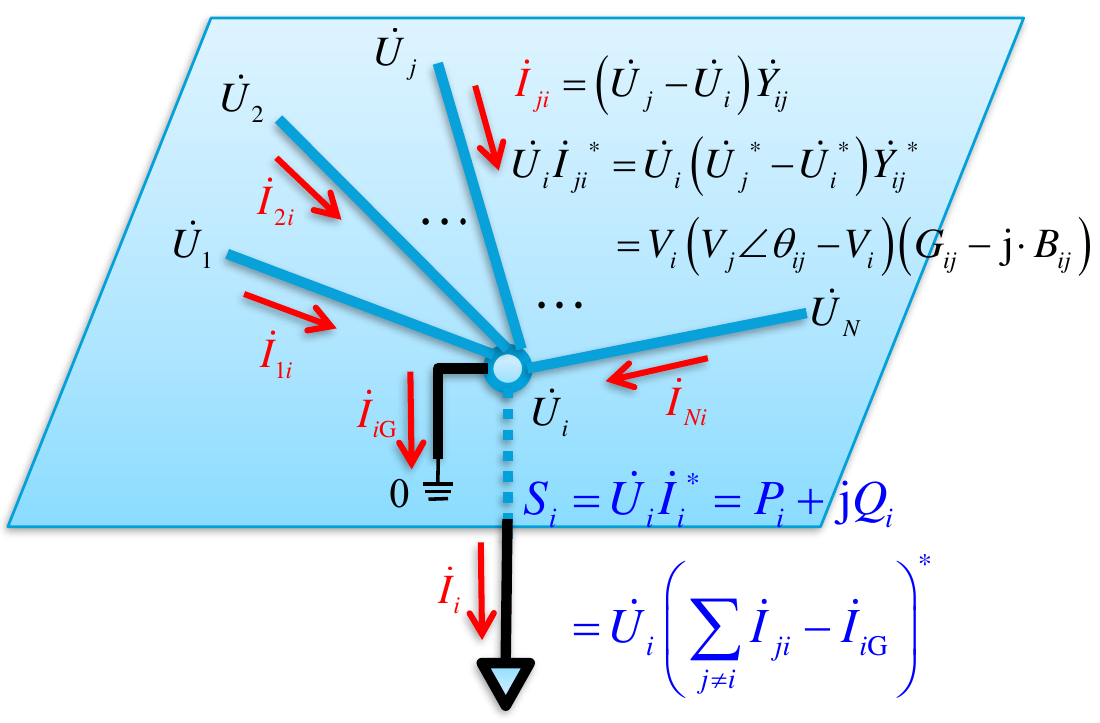}
\caption{Schematic Diagram  for Grid Network Operation}
\label{fig:GridOpr}
\end{figure}

Power flow (PF) analysis is a \emph{fundamental and most frequently used tool} for many tasks in a power system, such as fault diagnosis, state estimation, $N\!-\!1$ security assessment, and optimal power dispatch.
For each node $i$ in a grid network, choosing the reference direction as shown in Fig \ref{fig:GridOpr}, Kirchhoff's current law and ohm's law say that:
\begin{equation}
\label{eqI}
{{\dot{I}}_{i}}=\sum\limits_{\begin{smallmatrix}
 j=1 \\
 j\ne i
\end{smallmatrix}}^{n}{{{{\dot{I}}}_{j}}}=\sum\limits_{j\ne i}{{{{\dot{Y}}}_{ij}}\cdot \left( {{{\dot{U}}}_{j}}-{{{\dot{U}}}_{i}} \right)}.
\end{equation}
where ${\dot{Y}}_{ij}\!=\!G_{ij}\!+\!\mathrm{j} \cdot B_{ij}$  is the admittance in Cartesian form\footnote{$G$ is the conductance, $B$ is the susceptance, and $\mathrm{j}$ is the imaginary unit.}, and ${{\dot{U}}_{i}}\!=\!\left| {{{\dot{U}}}_{i}} \right|\angle {{\theta }_{i}}\!=\!{{V}_{i}}\angle {{\theta }_{i}}\!=\!{{V}_{i}}{{\text{e}}^{\text{j}{{\theta }_{i}}}}$ and ${{\dot{I}}_{i}}=\left| {{{\dot{I}}}_{i}} \right|\angle {{\phi }_{i}}$ are node voltage and node current, respectively, in polar form\footnote{${{V}_{i}}\angle {{\theta }_{i}}={{V}_{i}}{{\text{e}}^{\text{j}{{\theta }_{i}}}}={{V}_{i}}(\cos {{\theta }_{i}}+\text{j}\cdot \sin {{\theta }_{i}}).$}.

And thus, for each node in a power grid, Node $i$ for instance, taking account of the \textit{node-to-ground admittance} $y_i$, its active power $P$ and reactive power $Q$ are expressed as:
\begin{normalsize}
\begin{small}
\begin{equation}
\label{eq:PQend}
\begin{aligned}
  & \left\{ \begin{aligned}
  & {{P}_{i}}\!=\!{{V}_{i}}\sum\limits_{k\ne i}{{{V}_{k}}\left( {{G}_{ik}}\text{cos}{{\theta }_{ik}}\!+\!{{B}_{ik}}\text{sin}{{\theta }_{ik}} \right)}\!-\!{{V}_{i}}^{2}\sum\limits_{k\ne i}{{{G}_{ik}}}\!-\!{{V}_{i}}^{2}{{g}_{i}} \\
 & {{Q}_{i}}\!=\!{{V}_{i}}\sum\limits_{k\ne i}{{{V}_{k}}\left( {{G}_{ik}}\text{sin} {{\theta }_{ik}}\!-\!{{B}_{ik}}\text{cos}{{\theta }_{ik}} \right)}\!+\!{{V}_{i}}^{2}\sum\limits_{k\ne i}{{{B}_{ik}}}\!+\!{{V}_{i}}^{2}{{b}_{i}} \\
\end{aligned} \right.
\end{aligned}
\end{equation}
\end{small}
\end{normalsize}
Abstractly, a physical power system obeying Eq.~\eqref{eq:PQend} can be viewed as an analog engine---it takes bus voltage magnitude $V$ and bus voltage phaser $\theta$ as \emph{inputs}, conductance $G$ and susceptance $B$ as \emph{given parameters}, and ``computes"  active power injection $P$ and reactive power injection $Q$  as \emph{outputs}.

\subsection{Model-based Power Flow Analysis}
\begin{figure}[ht]
\centering
\subfloat[Conventional Model-based PF Analysis]{\label{fig:PF1}
\includegraphics[width=0.49\textwidth]{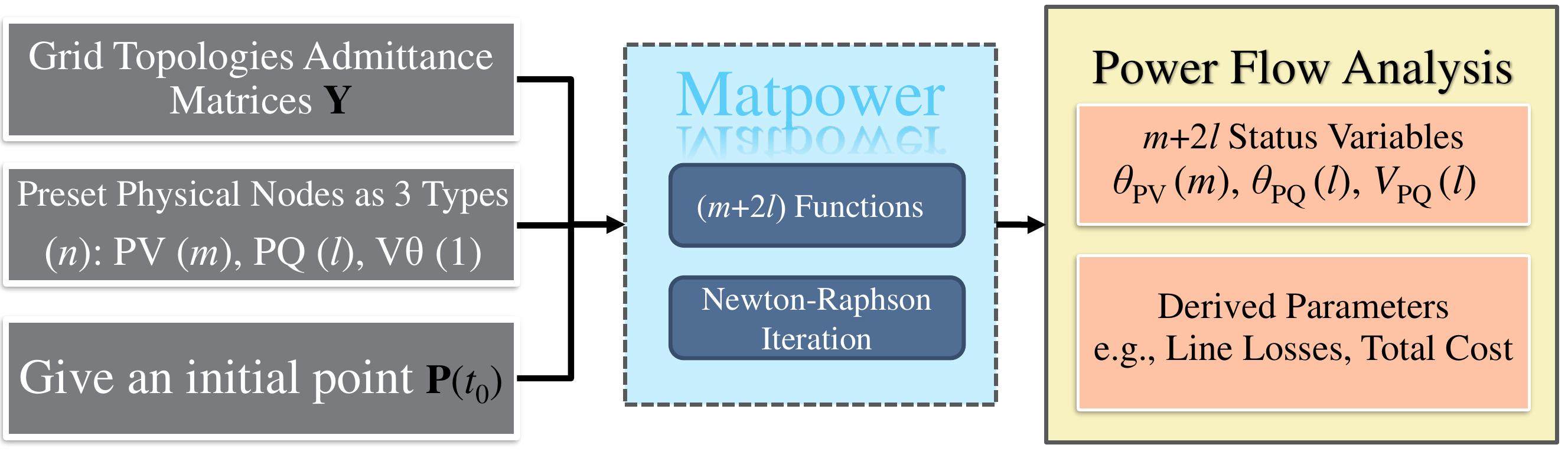}}

\subfloat[Data-driven PF Analysis]{\label{fig:PF2}
\includegraphics[width=0.20\textwidth]{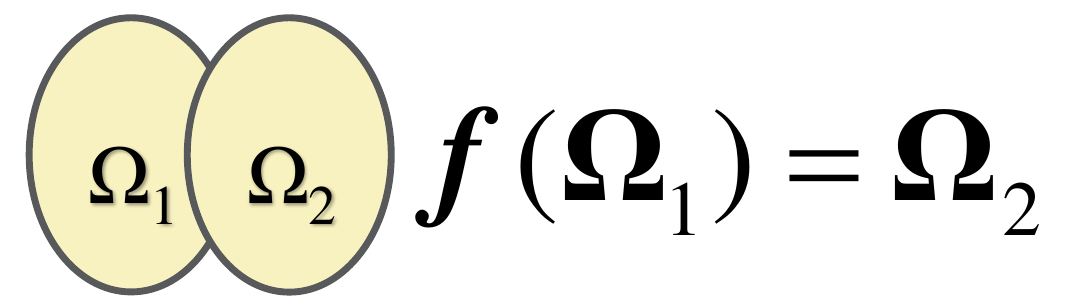}}

\caption{Model-based PF and Data-driven PF}
\label{fig:PFana}
\end{figure}

Before exploring PSDT for power flow SA, we revisit the conventional PF calculation. Firstly we give the technical roadmaps of both, as shown in Fig. \ref{fig:PFana}.

\subsubsection{Classical PF Formulation}
{\text{\\}}

Fig. \ref{fig:PF1} depicts a conventional PF calculation.
Its solution is model- and assumption-based. That is to say, the information of topological parameters, i.e., the admittance $Y$,  are \textit{prerequisite} for the calculation, and the input (output) variables \textit{need to be preset} as one of the following three categories:
\begin{itemize}
\item $P$ and $V$ ($Q$ and $\theta$) for voltage controlled bus, $PV$ bus;
\item $P$ and $Q$ ($V$ and $\theta$) for load bus, $PQ$ bus;
\item $V$ and $\theta$ ($P$ and $Q$) for reference bus, slack bus.
\end{itemize}

Conventional PF calculation deals mainly with the calculation of status variables, i.e., voltage magnitude $V$ and phase $\theta$, for each network bus, for a given set of variables such as load demands, i.e., active power $P$ and reactive power $Q$, under certain assumptions such as in a balanced steady-state system operation \cite{gomez2018electric}. With PF analysis, system operation conditions,  e.g., power flows and power losses of each line in the grid network, and reactive power outputs of the generators, can be determined \cite{vaccaro2018knowledge}.

Consider a grid with $n$ buses, among which there are  $m$ $PV$ buses,  $l$  $PQ$ buses, and $1$ slack bus $(n\!=\!m+\!l+\!1)$. Starting with Eq. \eqref{eq:PQend}, PF analysis is formulated as Eq.~\eqref{eq:J1}, which solves a set of equations with an equal number $(p\!=\!m+\!2l)$ of unknowns.

\begin{equation}
\label{eq:J1}
\mathbf{y}\!:=\! \left[ \begin{matrix}
   {{P}_{1}}  \\
   \vdots   \\
   {{P}_{n-1}}  \\
   {{Q}_{m+1}}  \\
   \vdots   \\
   {{Q}_{n-1}}  \\
\end{matrix} \right]\!=\!\bm f\left[ \begin{matrix}
   {{\theta }_{1}}  \\
   \vdots   \\
   {{\theta }_{n-1}}  \\
   {{V}_{m+1}}  \\
   \vdots   \\
   {{V}_{n-1}}  \\
\end{matrix} \right]\!=:\!\bm f\left( \mathbf{x} \right)
\qquad \mathbf{J}\!=\!\left[ \begin{matrix}
   \frac{\partial {{y}_{1}}}{\partial {{x}_{1}}} & \cdots  & \frac{\partial {{y}_{1}}}{\partial {{x}_{p}}}  \\
   \vdots  & \ddots  & \vdots   \\
   \frac{\partial {{y}_{p}}}{\partial {{x}_{1}}} & \cdots  & \frac{\partial {{y}_{p}}}{\partial {{x}_{p}}}  \\
\end{matrix} \right]
\end{equation}
where $:=$ is the assignment symbol in computer science.

Eq.~\eqref{eq:J1} builds a differentiable mapping $\bm f$, which consists of $p$ equations,  from the state variables, $\theta$ and $V$, to the power injections, $P$ and $Q$, i.e.,  $\bm f\!:\!\mathbf{x}\!\in\! {{\mathbb{R}}^{p}}\!\to\! \mathbf{y}\!\in\! {{\mathbb{R}}^{p}}$.

\subsubsection{Jacobian Matrix Estimation}
{\text{\\}}

Jacobian matrix $\mathbf J$  is a matrix of all first-order partial derivatives of a vector-valued function. For PF analysis, it is a sparse matrix that results from a sensitivity analysis of PF equations.
Tying together Eq.\eqref{eq:PQend} and Eq.\eqref{eq:J1}, we define the entries of $\mathbf J$, i.e., ${\left[ J \right]}_{ij}$, as the partial derivatives of the outputs, $P$ and $Q$, with respect to the inputs, $V$ and $\theta$. All in all, $\mathbf J$ consists of four parts  $\mathbf H, \mathbf N, \mathbf K, \mathbf L$ as follows:
\begin{normalsize}
\begin{small}
\begin{equation}
\label{eq:HNKLend}
\left\{ \begin{aligned}
  & {{H}_{ij}}\!={{V}_{i}}{{V}_{j}}\left( {{G}_{ij}}\sin {{\theta }_{ij}}\!-\!{{B}_{ij}}\cos {{\theta }_{ij}} \right)\!-\!{{\delta }_{ij}}\!\cdot\! {{Q}_{i}}\!+\!{{\delta }_{ij}}\!\cdot\! {V}_{i}^2 b_i \\
 & {{N}_{ij}}\!={{V}_{i}}{{V}_{j}}\left( {{G}_{ij}}\cos {{\theta }_{ij}}\!+\!{{B}_{ij}}\sin {{\theta }_{ij}} \right)\!+\!{{\delta }_{ij}}\!\cdot\! {{P}_{i}}\!-\!{{\delta }_{ij}}\!\cdot\! {V}_{i}^2 g_i \\
 & {{K}_{ij}}\!=-{{V}_{i}}{{V}_{j}}\left( {{G}_{ij}}\cos {{\theta }_{ij}}\!+\!{{B}_{ij}}\sin {{\theta }_{ij}} \right)\!+\!{{\delta }_{ij}}\!\cdot\! {{P}_{i}}\!+\!{{\delta }_{ij}}\!\cdot\! {V}_{i}^2 g_i \\
 & {{L}_{ij}}\!={{V}_{i}}{{V}_{j}}\left( {{G}_{ij}}\sin {{\theta }_{ij}}\!-\!{{B}_{ij}}\cos {{\theta }_{ij}} \right)\!+\!{{\delta }_{ij}}\!\cdot\! {{Q}_{i}}\!+\!{{\delta }_{ij}}\!\cdot\! {V}_{i}^2 b_i \\
\end{aligned} \right.
\end{equation}
\end{small}
\end{normalsize}
where $ {{H}_{ij}}\!=\!\frac{\partial {{P}_{i}}}{\partial {{\theta }_{j}}}, {{N}_{ij}}\!=\!\frac{\partial {{P}_{i}}}{\partial {{V}_{j}}}{{V}_{j}}, {{K}_{ij}}\!=\!\frac{\partial {{Q}_{i}}}{\partial {{\theta }_{j}}}, {{L}_{ij}}\!=\!\frac{\partial {{Q}_{i}}}{\partial {{V}_{j}}}{{V}_{j}}$.

And then
\[
\mathbf{J}= \left[ \begin{array}{*{35}{l}}
   {{\left[ \mathbf H \right]}_{n-1, n-1}} & {{\left[ \mathbf N \right]}_{n-1, l}}  \\
   {{\left[ \mathbf K \right]}_{l, n-1}} & {{\left[ \mathbf L \right]}_{l, l}}  \\
\end{array} \right]
\]

$\mathbf {J}$ inherently contains the information about the most up-to-date network topology, which can be taken care of by the topology processing in the Energy Management System.
In practice, however, \textit{$\mathbf {J}$ may not be accurately obtained} due to following reasons: 1)  topology error has long been cited as a major cause of inaccurate estimation results \cite{Wu1989Detection}: network topologies may be out-of-data due to erroneous record, delay telemetry, or unexpected operation, \textit{especially for a distribution network}, and line impedance parameters may be susceptible to climate changing and usage lifetime; 2) uncertainty and individuality of customer units, which are small in size but large in amount \cite{he2019invisible}; 3) ubiquitous noises, e.g., load/DG fluctuations, and 4) inevitable measurement errors, such as data missing, abnormal, and out of sync.

\subsubsection{PF Analysis based on Jacobian Matrix Estimation}
{\text{\\}}

Under the model-based mode, these two problems, i.e., PF calculation and $\mathbf J$ estimation, are closely intertwined---the determination of  $\mathbf J$ is an essential part for PF calculation \cite{vaccaro2018knowledge}.
This phenomena introduces an additional uncertainty for the PF analysis, and also \textit{sets a high access threshold for the analysis}.
Numerical iteration algorithms and sparse factorization techniques, mainly based on Newton-Raphson and fast-decoupled methods, are used to approximate the \textit{nonlinear PF equations} by linearized  $\mathbf J$ \cite{7364281}.
%
%

To formulate the linear approximation process that the system operation point shifts from $(\mathbf{x}^{(k)},\mathbf{y}^{(k)})$ to $(\mathbf{x}^{(k\!+\!1)},\mathbf{y}^{(k\!+\!1)})$, the iteration is set as follows:
\begin{equation}
\label{eq:XJ}
{{\mathbf{x}}^{\left( k+1 \right)}}:={{\mathbf{x}}^{\left( k \right)}}+{\mathbf {J}}^{-1}\left( {{\mathbf{x}}^{\left( k \right)}} \right)\left( {{\mathbf{y}}^{\left( k+1 \right)}}-{{\mathbf{y}}^{\left( k \right)}} \right)
\end{equation}

The iteration, given in Eq.  \eqref{eq:XJ}, depicts how to update the state variables from ${\mathbf{x}}^{\left( k \right)}$ to ${\mathbf{x}}^{\left( k+1 \right)}$.
${\mathbf{y}}^{\left( k \right)}$ and ${\mathbf{x}}^{\left( k \right)}$ are known variables which are measurable or calculable.
${\mathbf{y}}^{\left( k+1 \right)}$, according to Eq.~\eqref{eq:J1}, is our desired $P, Q$ on $PQ$ buses and desired $P$ on $PV$ buses\footnote{For $PQ$ buses, neither $V$ nor $\theta$ are fixed; they are state variables that need to be estimated.  For $PV$ buses, $V$ is fixed, and $\theta$ needs to be estimated.}.
Then we focus on  $\mathbf {J}$  for the iteration.
Traditionally, $\mathbf {J}$ is computed via Eq. \eqref{eq:HNKLend}.
The model-based approach, however, is not ideal in practice, since the up-to-date network topology and relevant parameters (admittance $Y$), and the operation points $(\mathbf{x}^{(k)},\mathbf{y}^{(k)})$ are required \emph{at to be of high resolution and high precision}; these requirements, unfortunately, are often unrealistic as aforementiond.

\section{DT for Real-time Power Flow Analysis}
\subsection{Background of the Cases}
\label{Sec:IEEE9case}
Cases are built upon the simulation tool Matpower \cite{he2015arch}. Specifying the power injection on each node ($\mathbf y$ in Eq. \eqref{eq:J1}), we solve the PF equation (Eq. \eqref{eq:PQend}) with the known network model (given parameters $G$ and $B$), and obtain the voltage magnitude and angle ($\mathbf x$ in Eq. \eqref{eq:J1}).

For a standard IEEE 9-bus system, Node 1 is the slack bus, Node 2, 3 are the PV buses, Node 5, 7, 9 are the PQ buses with actual load injection, and Node 4, 6, 8 are the PQ buses without load or generator injection, also seemed as tie line.

\begin{figure}[http]
\centering
\includegraphics[width=0.46\textwidth]{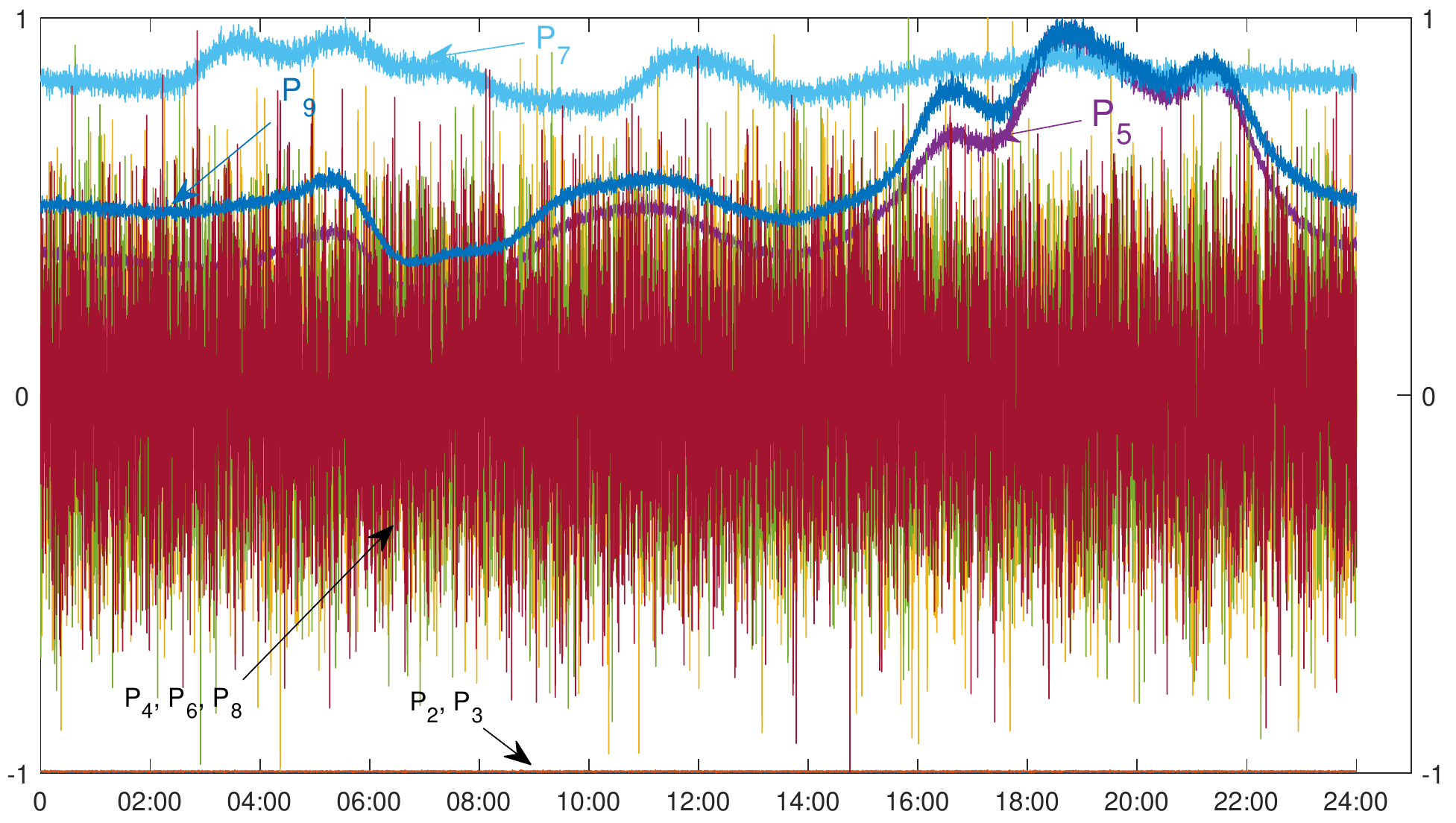}
\caption{Power Consumption for IEEE 9-Bus System}
\label{fig:dailyP}
\end{figure}

Suppose the sampling dataset is with 9600 points\footnote{The sampling rate for PMU, according to IEC-61850 standard, can reach up to 4800 Hz \cite{sun2017dynamic}, and for micro-PMU ($\mu$PMU) can reach 120 Hz \cite{mohsenian2018distribution}.}.
Considering the \textit{i.i.d. Gaussian power fluctuations} according to reference \cite{he2015arch}, the normalization value of active power $P$ is obtained as shown in Fig.~\ref{fig:dailyP}.  The  reactive power $Q$ has similar trend. Note that the raw values of $P$ and $Q$ on Node 5, 7, 9 (with actual load injection) are much larger than that on Node 4, 6, 8 (tie line). Thus, the standardization processes, consisting of an addition of very small artificial noise and the normalization of the matrix, do cause much larger amplitude vibrations on tie line Node 4, 6, 8.

\subsection{DT for Real-time Power Flow Monitoring}
This section explores DT based on Artificial Neural Network (ANN), a typical AI algorithm as stated in Sec. \ref{sec:BDAandAI}.
ANN involves a network of simple processing elements (artificial neurons) which can exhibit complex global behavior, determined by connections between the processing elements and the element parameters. In most cases ANN is an adaptive system that changes its structure based on external or internal information that flows through the network.

Following the roadmaps given in Fig. \ref{fig:PF2}, we build a 5 layers ANN to map the \textit{non-linear} relationship $\bm f$ between the outputs ($\mathbf y$ in  Eq. \eqref{eq:J1}, $P, Q$)  and the inputs ($\mathbf x$ in Eq. \eqref{eq:J1}, $\theta, V$), i.e., $\mathbf{P}=\bm f\left( \mathbf{V},\bm{\theta } \right).$
The neural number for each layers is set as $[14, 50, 50, 50, 14]$, and tanh is chosen as the activate function.

\begin{figure}[ht]
\centering
\subfloat[Prediction of P on load nodes]{\label{fig:NN1}
\includegraphics[width=0.24\textwidth]{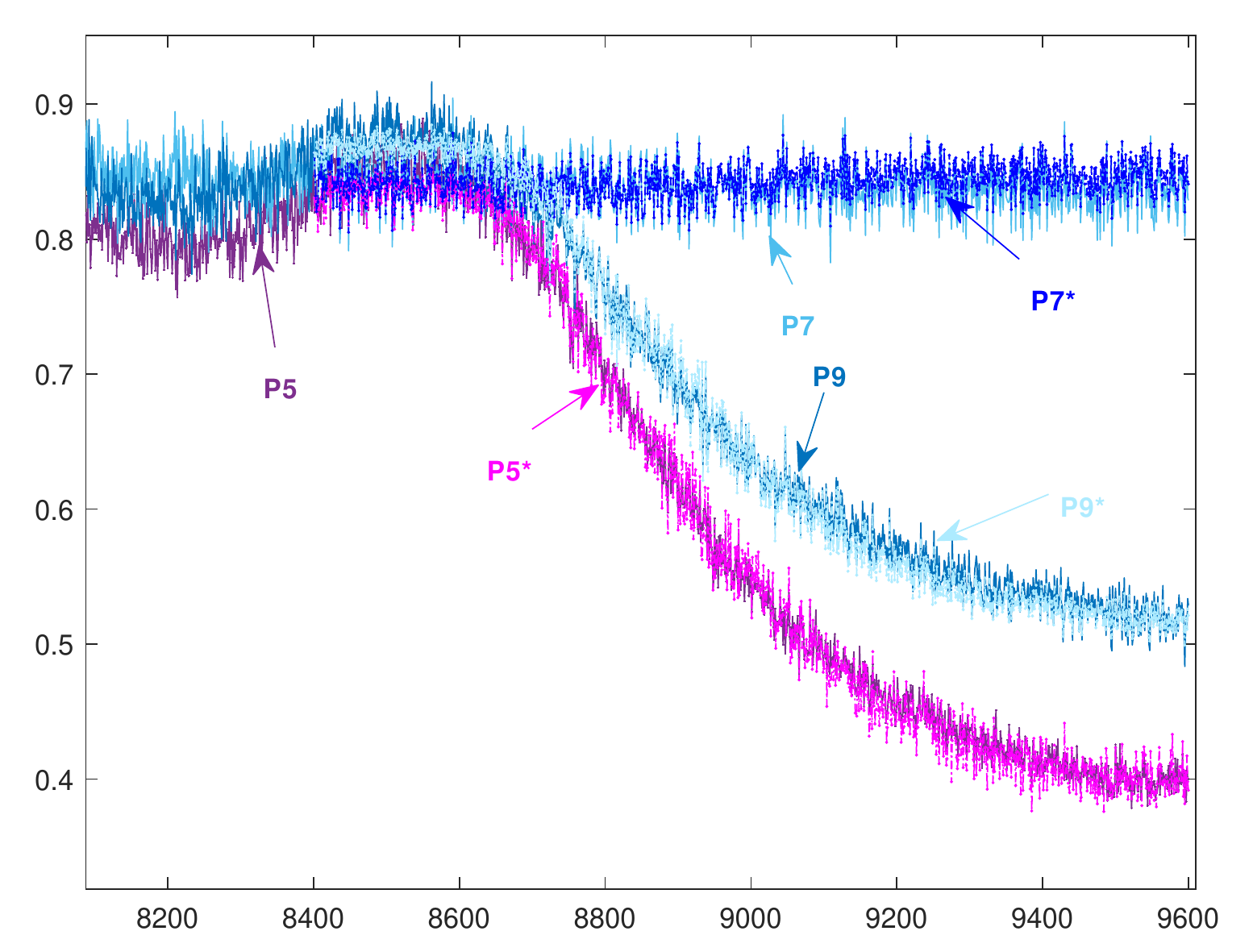}
}
\subfloat[$\mathbf J$ estimation via Eq. \eqref{eq:JL}]{\label{fig:NN2}
\includegraphics[width=0.22\textwidth]{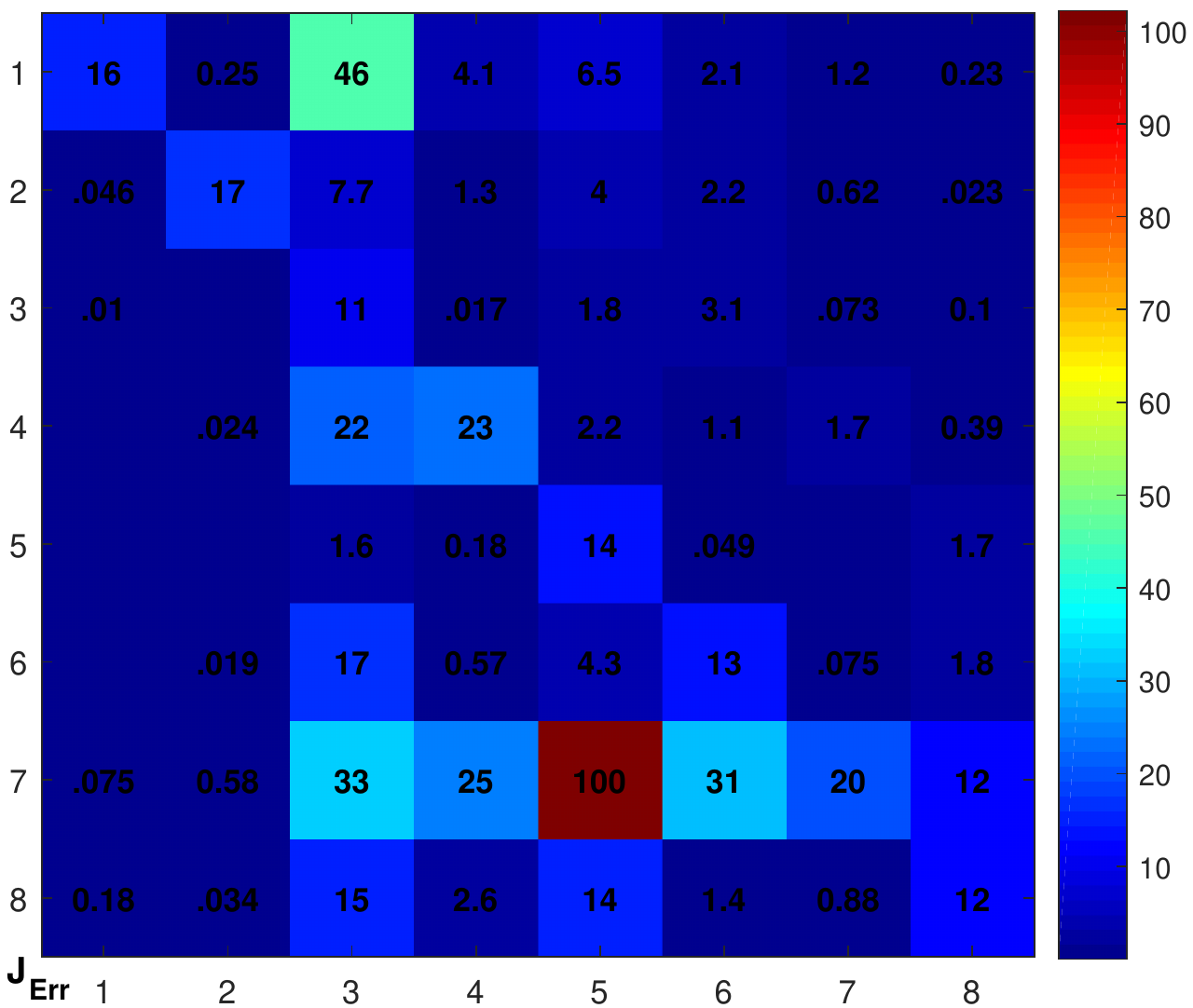}
}
\caption{Power Prediction and $\mathbf J$ Estimation Using ANN}
\label{fig:NN}
\end{figure}

We use the data during 1:8400 for training (seen as a regression problem, and the label is $\mathbf y$), and during 8401:9600 for testing. Taking the active power $P$ from the PQ buses with actual load injection (i.e., Node 5, 7 and 9) as the regression targets,  Fig. \ref{fig:NN1} shows the result---the regression value $P_5^*, P_7^*, P_9^*$ are very close to the truth-value $P_5, P_7, P_9$.

This PF Monitoring DT is \textit{data-driven} and \textit{real-time}. A quite well performance is achieved with a very simple start---only operation data are needed. PSDT is capable of handling the scenarios where system topologies and network parameters are \textit{unreliable or even totally unavailable}, and thus, the physical models and admittance $Y$ are no longer required information.
Moreover, this PF monitoring DT is independent from Jacobian matrix $\mathbf J$.

\subsection{DT for Real-time Jacobian Matrix Estimation}
\subsubsection{Background and Benchmark for $\mathbf J$ Estimation}
{\text{\\}}

\begin{figure}[ht]
\centering
\subfloat[Truth-value of  $\mathbf J_0$]{\label{fig:Jmean}
\includegraphics[width=0.25\textwidth]{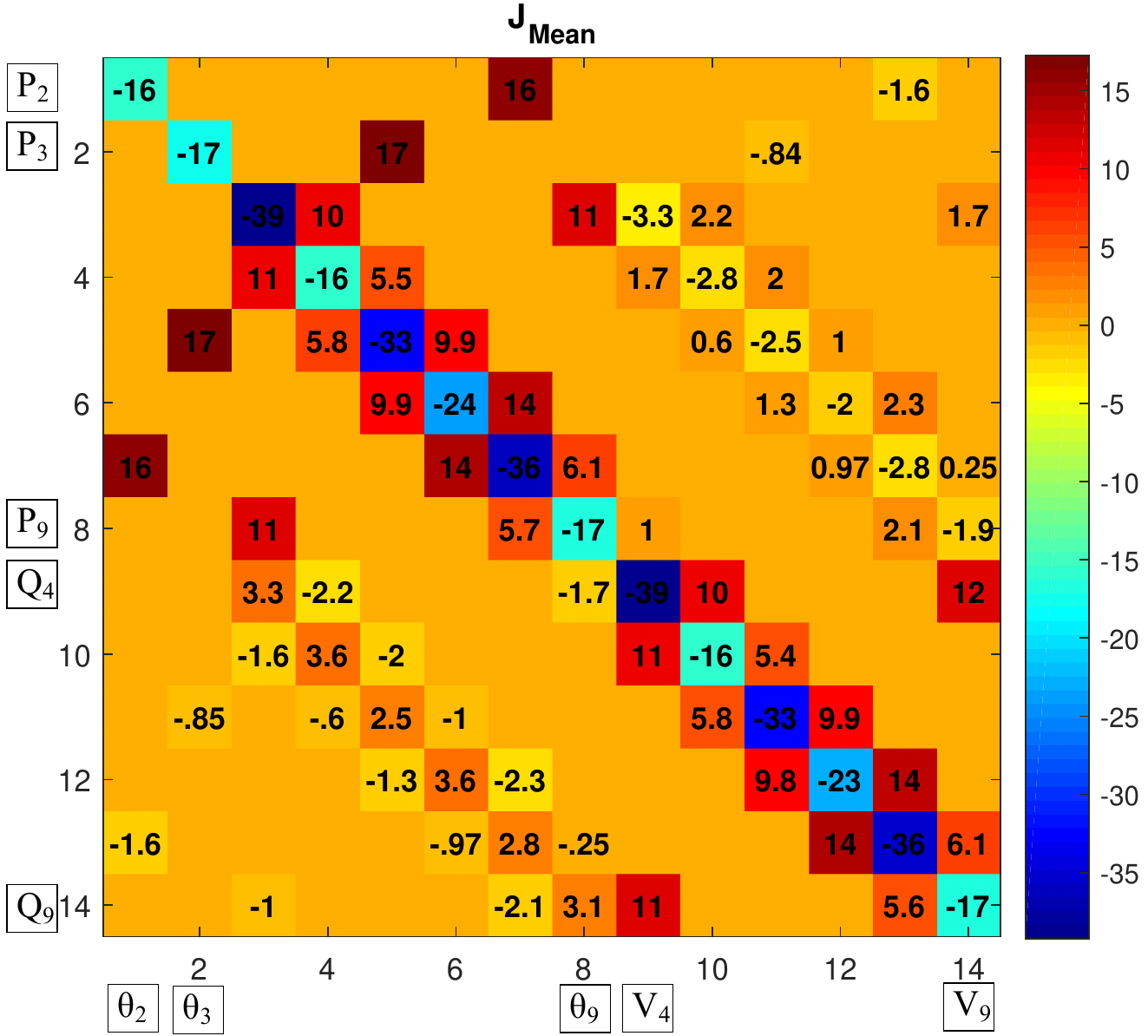}}
\subfloat[Standard deviation of $\mathbf J_0$]{\label{fig:Jstd}
\includegraphics[width=0.23\textwidth]{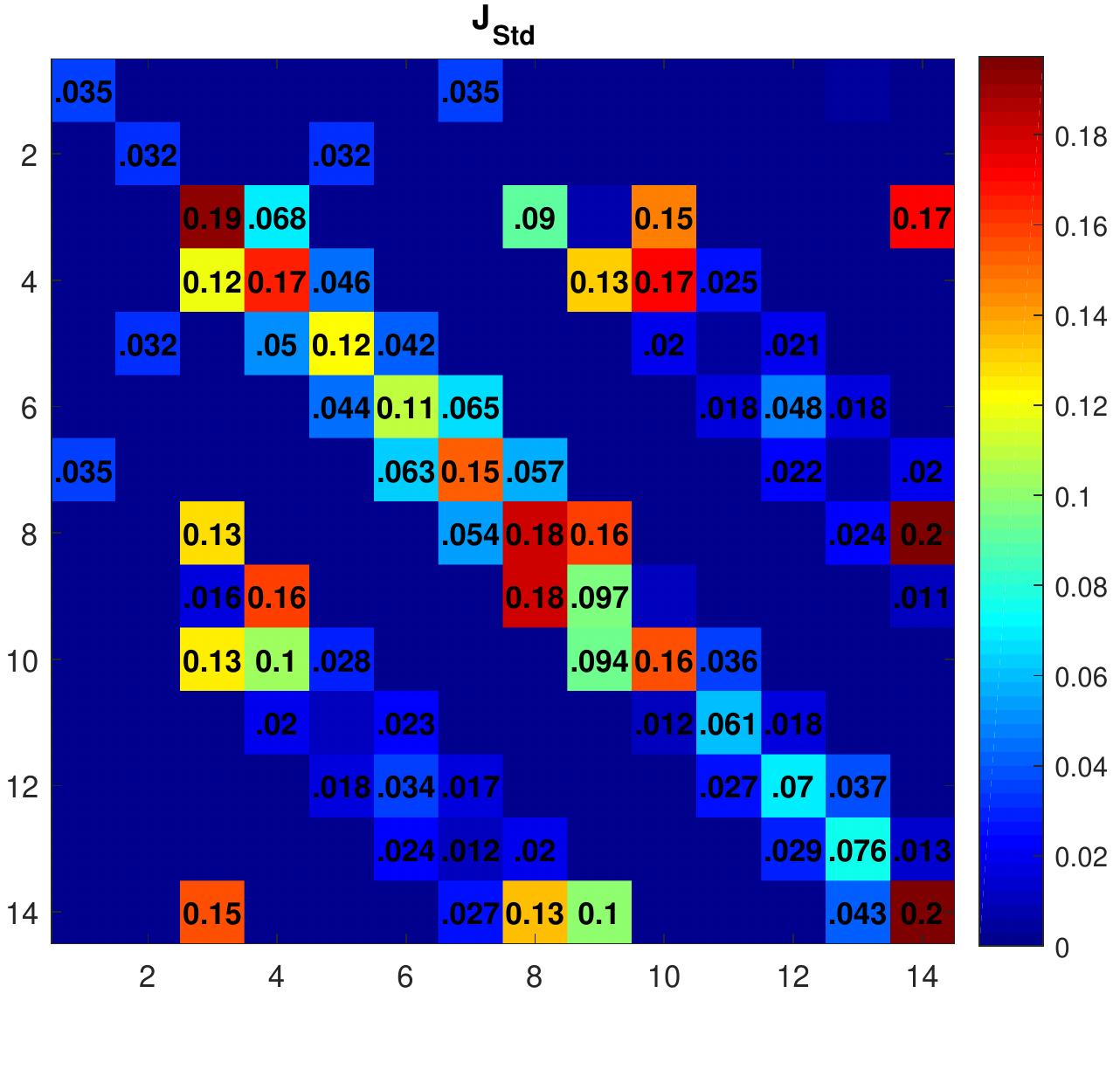}}
\caption{Basic Statistical Information of $\mathbf J$ from 9600 Samplings}
\label{fig:dailyJ0}
\end{figure}

Under \textit{fairly general conditions},  Jacobian matrix $\mathbf {J}$, according to Eq. \eqref{eq:HNKLend},  \textit{keeps nearly constant within a short time}, called $\Delta t$, due to the stability of the system, or concretely, of variables $V, \theta, Y$.
The truth-value of $\mathbf J$ is calculated via  Eq.~\eqref{eq:HNKLend} in a model-based way. During the 9600 observations (Fig. \ref{fig:dailyP}),  $\mathbf J$ keeps nearly constant at around $\mathbf J_{\text{Mean}}$ (Fig.~\ref{fig:Jmean}), and with a small standard deviation $\mathbf J_{\text{Std}}$ (Fig.~\ref{fig:Jstd}). $\mathbf J_{\text{Mean}}$ is set as the benchmark during the whole period.

\subsubsection{SA---$\mathbf J$ Estimation with ANN}
{\text{\\}}

Naturally, we try to continue our $\mathbf J$ estimation task based on our well-trained ANN during our last real-time PF monitoring task.
The non-linear ANN is modeled as
\begin{normalsize}
\begin{small}
\[  \mathbf{y}\!=\!\bm f \left( \mathbf{x} \right)\! \triangleq \!{{\bm f}^{L}}\left( {{\mathbf{W}}^{L}}  \!\cdots\! {{\bm f}^{2}}\left( {{\mathbf{W}}^{\text{2}}}{{\bm f}^{1}}\left( {{\mathbf{W}}^{1}}\mathbf{x}\!+\!{{\mathbf{b}}^{1}} \right)\!+\!{{\mathbf{b}}^{\text{2}}} \right)\!\cdots\! \!+\!{{\mathbf{b}}^{L}} \right)\]
\end{small}
\end{normalsize}

With repeatedly use of the Chain Rule,  $\mathbf{J}$ is solved as
\begin{equation}
\label{eq:JL}
\begin{aligned}
  \mathbf{J} &=\frac{\partial \mathbf{y}}{\partial {{\mathbf{x}}^{\text{T}}}}\!=\!\frac{\partial {{\mathbf{a}}^{\left( L \right)}}}{\partial {{\mathbf{x}}^{\text{T}}}}\!=\!
  =\!\frac{\partial {{\bm f}^{\left( L \right)}}\left( {{\mathbf{z}}^{\left( L \right)}} \right)}{\partial {{\mathbf{z}}^{\left( L \right)}}^{\text{T}}}\frac{\partial \left( {{\mathbf{W}}^{L-1}}{{\mathbf{a}}^{L-1}}+{{\mathbf{b}}^{L-1}} \right)}{\partial {{\mathbf{x}}^{\text{T}}}} \\
 & =\!\text{diag}\left(\bm f^{ L '}_ {\mathbf z\!=\!{\mathbf{z}}^{ L }}\right) {{\mathbf{W}}^{L-1}}\frac{\partial {{\mathbf{a}}^{L-1}}}{\partial {{\mathbf{x}}^{\text{T}}}}\!  \triangleq \!{{\mathbf{\Gamma }}^{L}}{{\mathbf{W}}^{L-1}}\frac{\partial {{\mathbf{a}}^{L-1}}}{\partial {{\mathbf{x}}^{\text{T}}}}\\
  & ={{\mathbf{\Gamma }}^{L}}{{\mathbf{W}}^{L-1}}{{\mathbf{\Gamma }}^{L-1}}{{\mathbf{W}}^{L-2}}\cdots {{\mathbf{\Gamma }}^{2}}{{\mathbf{W}}^{1}} \\
\end{aligned}
\end{equation}
where ${{\mathbf{\Gamma }}^{l}}\!=\!\text{diag}\left(\bm f^{ l '}_ {\mathbf z\!=\!{\mathbf{z}}^{ l }}\right), l\!=\!2,\cdots,L$

Following Eq.~\eqref{eq:JL}, we obtain the result (Fig. \ref{fig:NN2}), and find the task fail.
It can deduce that \textit{the direct use of ANN may be unsuitable for handling derivative signal analysis, although a quite good result could be obtained for the regression.} The derivative signal may have some connection with a residual network, and this topic will be discussed elsewhere.

\subsubsection{SA---$\mathbf J$ Estimation with Least-Square Estimation}
{\text{\\}}

$\mathbf {J}$ estimation is an inverse PF (IPF) problem \cite{yuan2016inverse}.
During some period, $t_{\Delta}$ for instance, considering $T$ observations at time instants $t_i$, $(i\!=\!1,2,\cdots,T, t_T\!-\!t_1\!=\!t_{\Delta})$, operation points $(\mathbf{x}^{(i)},\mathbf{y}^{(i)})$ are obtained in the form of  Eq. \eqref{eq:J1}. Defining $\Delta {{\mathbf{x}}^{\left( k \right)}}\!\triangleq\! {{\mathbf{x}}^{\left( k+1 \right)}}\!-\!{{\mathbf{x}}^{\left( k \right)}}$ and $\Delta {{\mathbf{y}}^{\left( k \right)}}\!\triangleq\! {{\mathbf{y}}^{\left( k+1 \right)}}\!-\!{{\mathbf{y}}^{\left( k \right)}}$, from Eq.  \eqref{eq:XJ} we deduce that $\Delta {{\mathbf{y}}^{\left( k \right)}}\!\approx\!\mathbf {J}^{\left( k \right)}\Delta {{\mathbf{x}}^{\left( k \right)}}$. Since $\mathbf {J}$ is nearly constant, the matrix form is formulated as
\begin{equation}
\label{Eq:MMYX}
\mathbf B \!\approx\! \mathbf J\mathbf{A}
\end{equation}
where $\mathbf{J} \!\in\! {{\mathbb{R}}^{N\!\times\!N}} $ ($N\!=\!p\!=\!m+2l$), $\mathbf{B} \!=\! \left[ {\Delta {{{\mathbf{y}}^{(1)}}}, \cdots ,\Delta  {{{\mathbf{y}}^{(T)}}}} \right] \!\in\! {{\mathbb{R}}^{N\!\times\!T}}$, and $\mathbf{A}\! =\! \left[ {\Delta {{{\mathbf{x}}^{(1)}}}, \cdots ,\Delta  {{{\mathbf{x}}^{(T)}}}} \right] \!\in\! {{\mathbb{R}}^{N\!\times\!T}}$.
Thus, we turn the estimation of $\mathbf{J}$  into a standard \textit{regression problem}, and the least-square estimation (LSE) is the first and most obvious choice as the solution\cite{7364281}. Rewriting  Eq. \eqref{Eq:MMYX} as
\begin{equation}
\label{Eq:LSENor}
\bm{\Theta} \approx \bm{\Lambda }\mathbf{J}^{\text{T}},
\end{equation}
where  $\bm{\Theta}\!:=\!\mathbf{B}^{\text{T}}\!\in\! {{\mathbb{R}}^{T\!\times\!N}}$ and  $\bm{\Lambda}\!:=\!\mathbf{A}^{\text{T}}\!\in\! {{\mathbb{R}}^{T\!\times\!N}}.$
Since $\bm{\Lambda}$ is \emph{over-determined}, i.e., $T\!>\!N$, the LSE says that
\begin{equation}
\label{Eq:LSEsol}
 {\hat{\mathbf{J}}^{\text{T}}}={{\left( {\bm{\Lambda}}^{\text{T}}\bm{\Lambda} \right)}^{-1}}{\bm{\Lambda}}^{\text{T}}\bm{\Theta}={\left( \mathbf{A}\mathbf{A}^{\text{T}} \right)}^{-1}\mathbf{A}\mathbf{B}^{\text{T}}
\end{equation}

Due to the ubiquitous power fluctuations and measurement errors, the matrix $\bm{\Lambda}$ \textit{would not be ill-conditioned under the aforementioned normal operation conditions}. This property guarantees the performance of LSE estimation.
Moreover, some small artificial noise can be added, according to our previous work \cite{he2015arch}, to prevent the (normalized) matrix $\bm{\Lambda}$ from being ill-conditioned.

\begin{figure}[tb]
\centering
\subfloat[Bias using 4800 samplings]{
\includegraphics[width=0.24\textwidth]{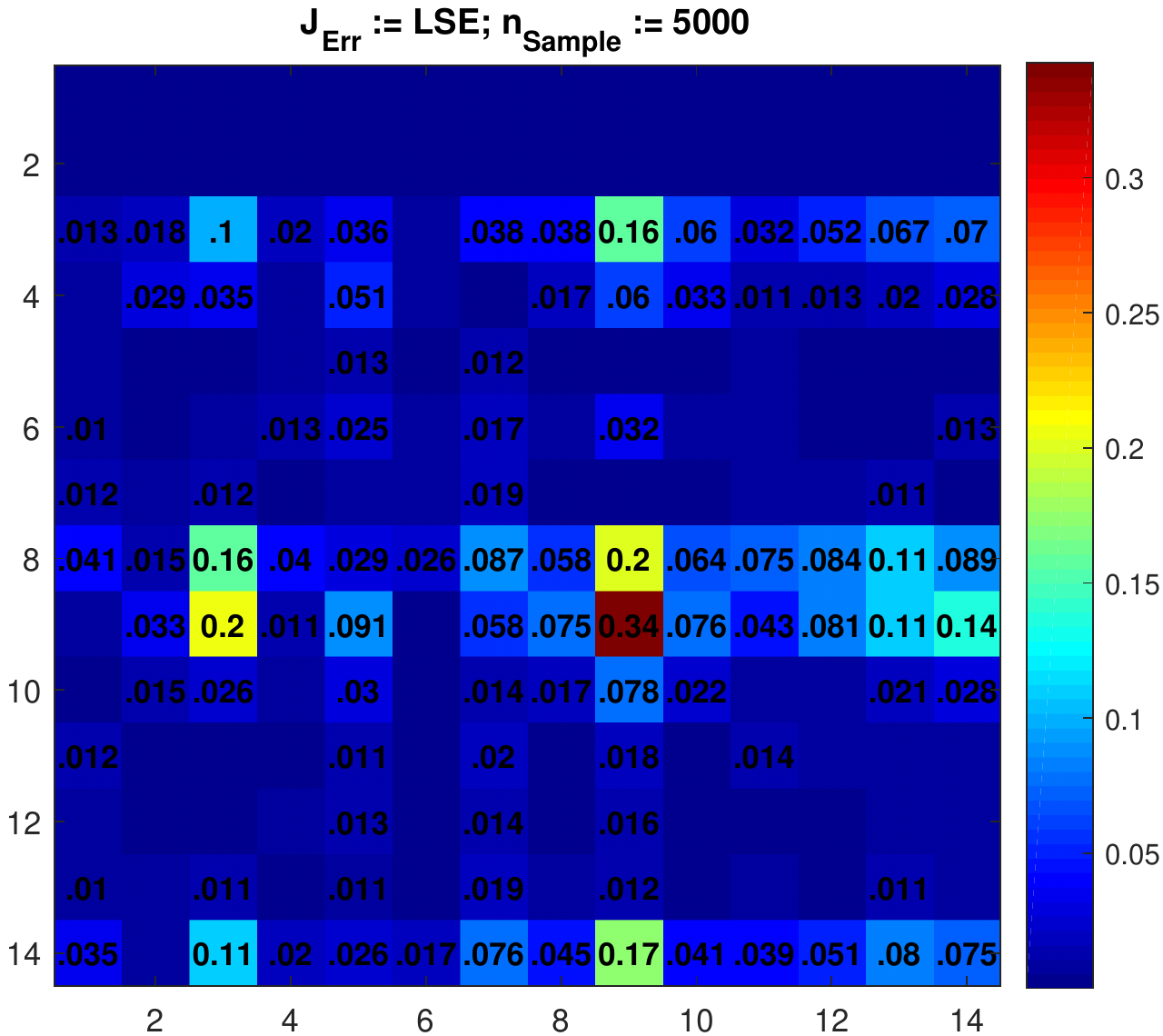}
}
\subfloat[Bias using 240 samplings]{
\includegraphics[width=0.24\textwidth]{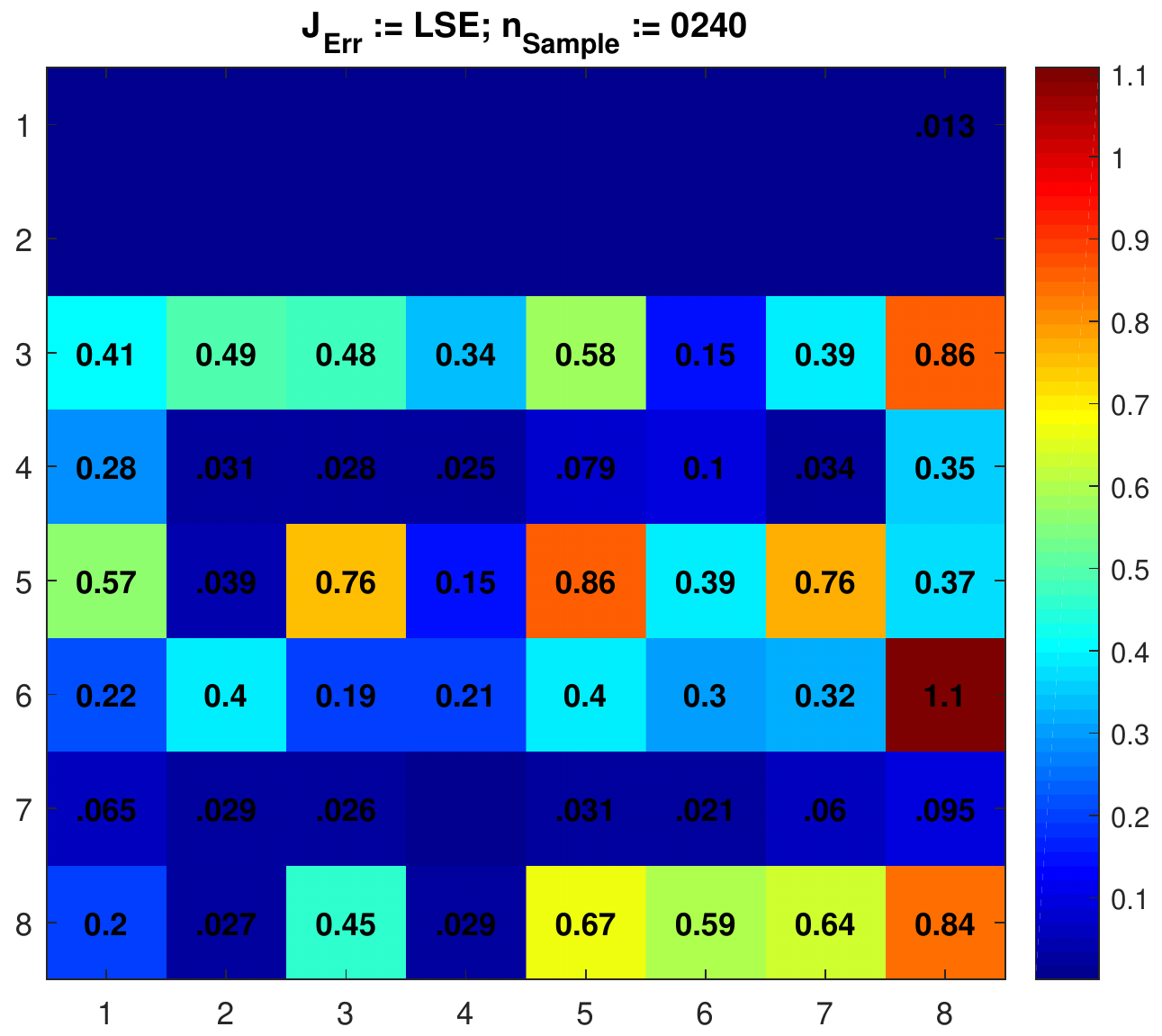}
}
\caption{Estimation Bias of $\mathbf J$ with Large and Small Dataset}
\label{fig:J5000}
\end{figure}

We use the proposed LSE to handle the large dataset and small dataset, respectively, and then compare the results with the benchmark (Fig.~\ref{fig:Jmean}). Fig. \ref{fig:J5000} gives the estimation bias.

It is observed that LSE has good performances on the $\mathbf J$ estimation task with a large dataset.
With a small dataset, the performances become worse.
Fig  \ref{fig:J5000} reveals that for the proposed data-driven $\mathbf J$ estimation algorithm,
\textit{increasing data collection will improve the performance}; it is not true, however, for model-based one.
Besides, data-driven $\mathbf J$ estimation no longer needs the admittance $\mathbf Y$. Conversely, the result of $\mathbf {J}$ estimation inherently contains information about the most up-to-date network topologies and corresponding parameters.

\subsubsection{Virtual Test---Closed-loop Feedback on $\mathbf J$ Estimation}
\label{sec:vtjac}
{\text{\\}}

\begin{figure}[ht]
\centering
\subfloat[Truth-value of  $\mathbf J_0$]{\label{fig:J1180}
\includegraphics[width=0.24\textwidth]{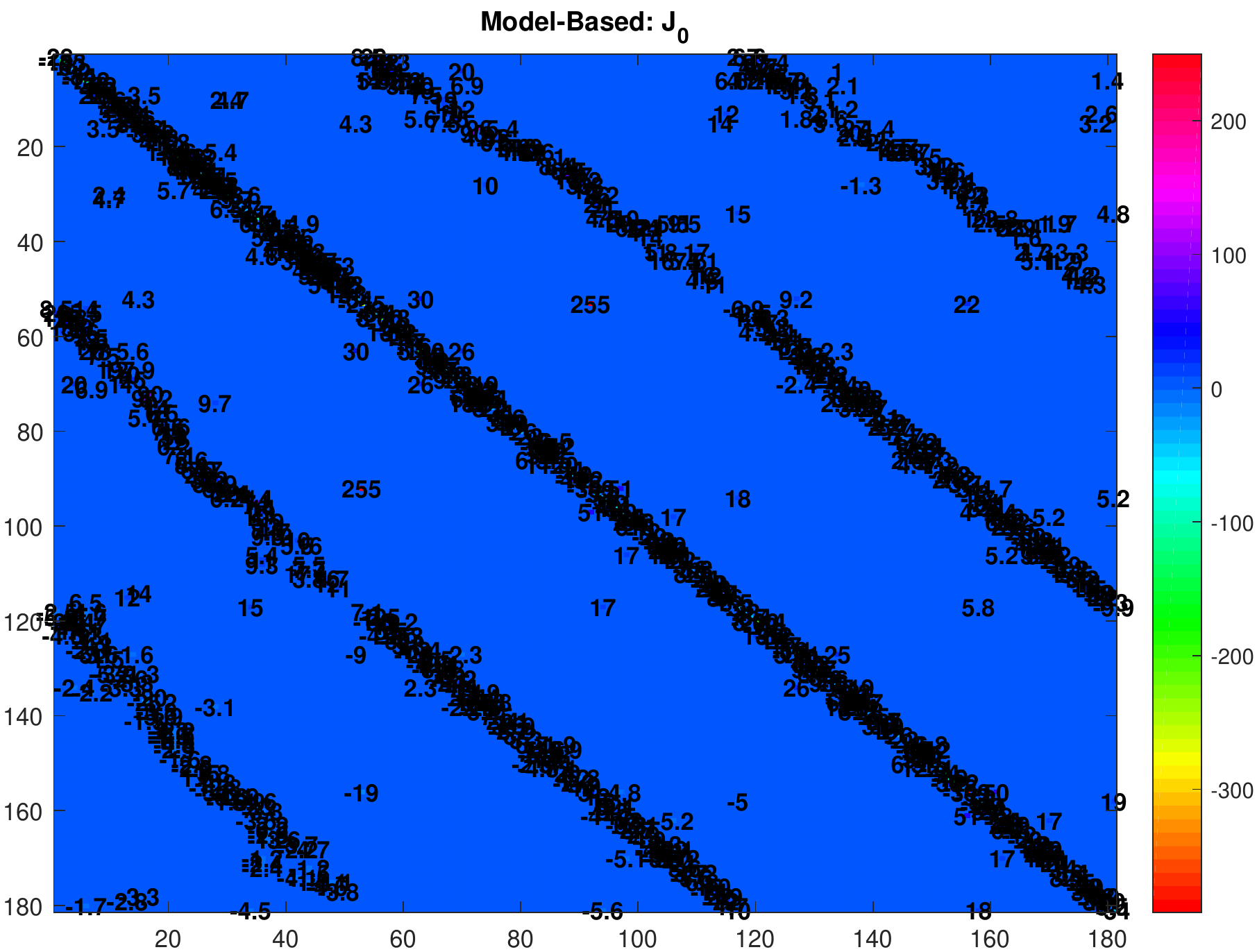}}
\subfloat[Bias of Estimation 1]{\label{fig:J1181}
\includegraphics[width=0.24\textwidth]{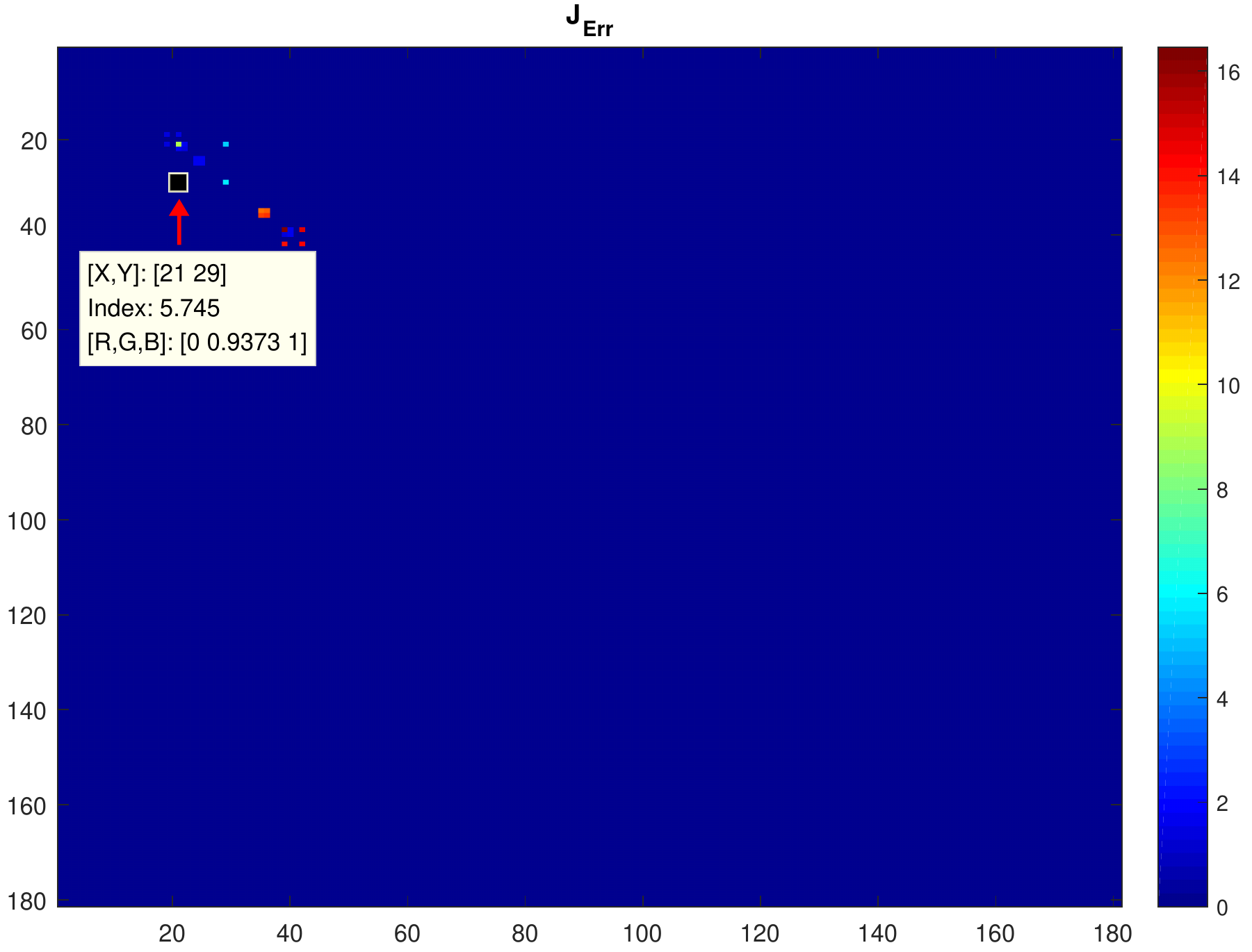}}

\subfloat[Bias of Estimation 2]{\label{fig:J1182}
\includegraphics[width=0.24\textwidth]{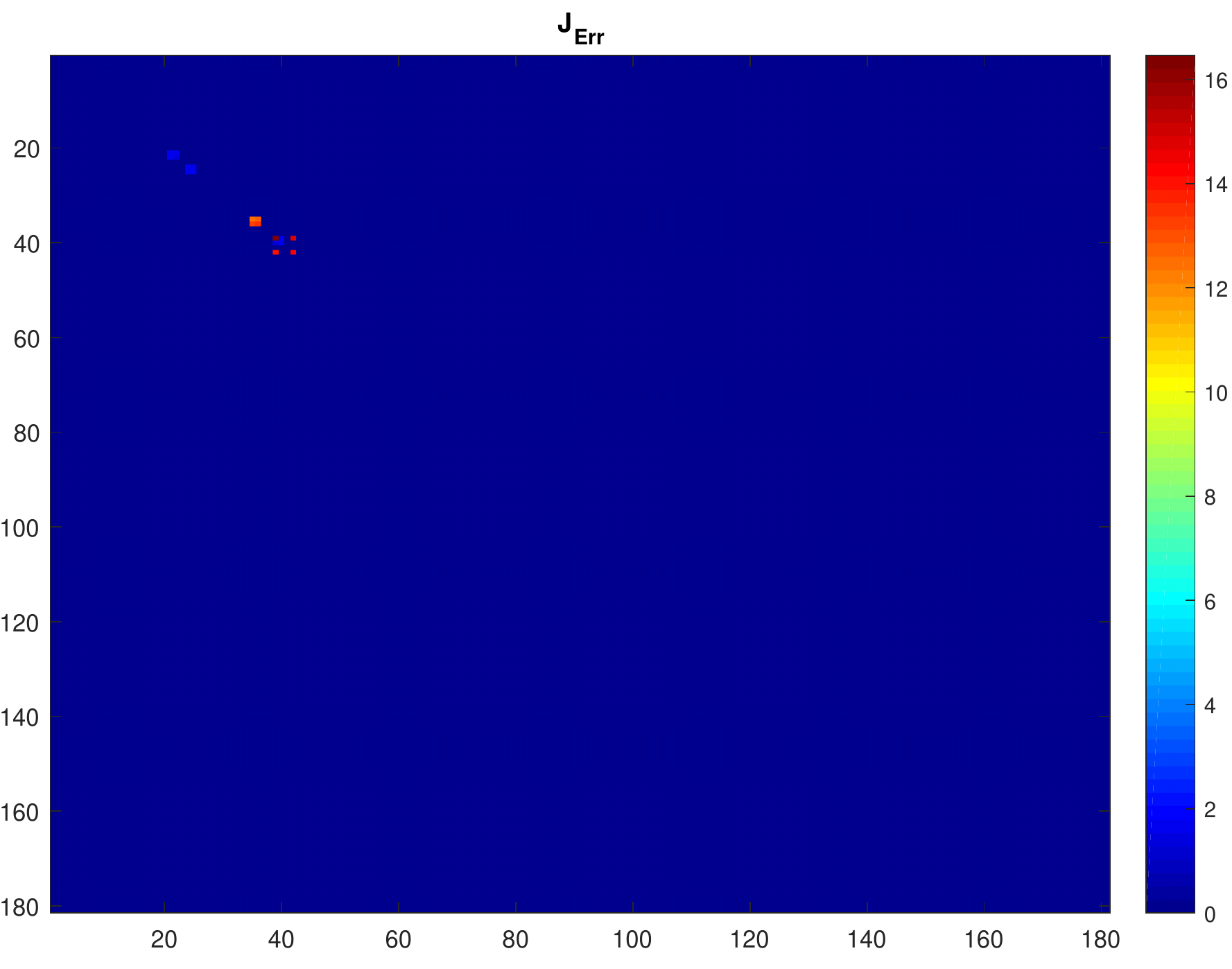}}
\subfloat[Bias of Estimation 3]{\label{fig:J1183}
\includegraphics[width=0.24\textwidth]{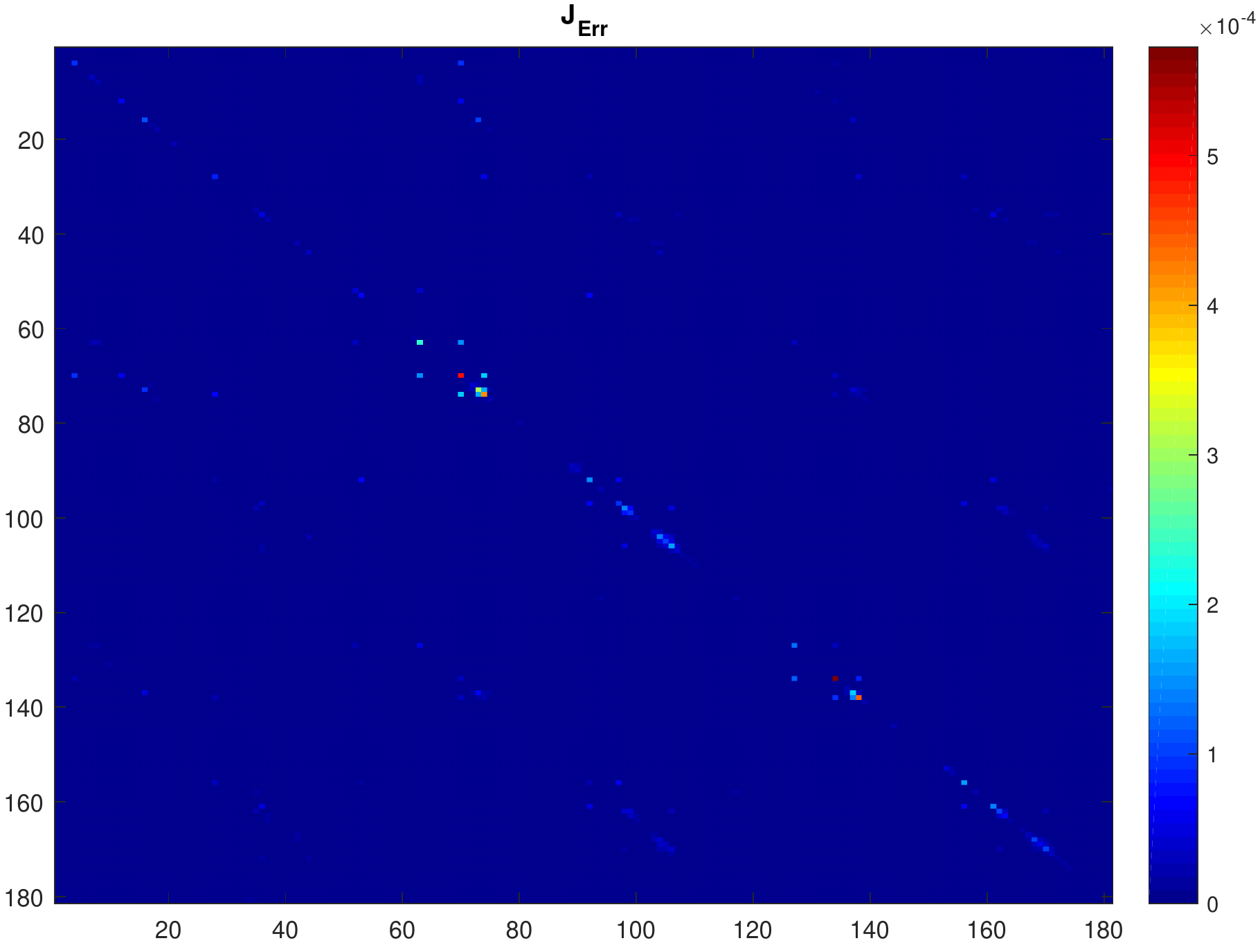}}

\subfloat[Description File for Branch]{\label{fig:bb1}
\includegraphics[width=0.48\textwidth, height=0.08\textheight]{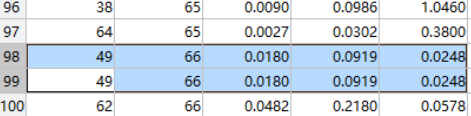}}

\subfloat[Factor Model for Bias 1]{\label{fig:fm1}
\includegraphics[width=0.24\textwidth]{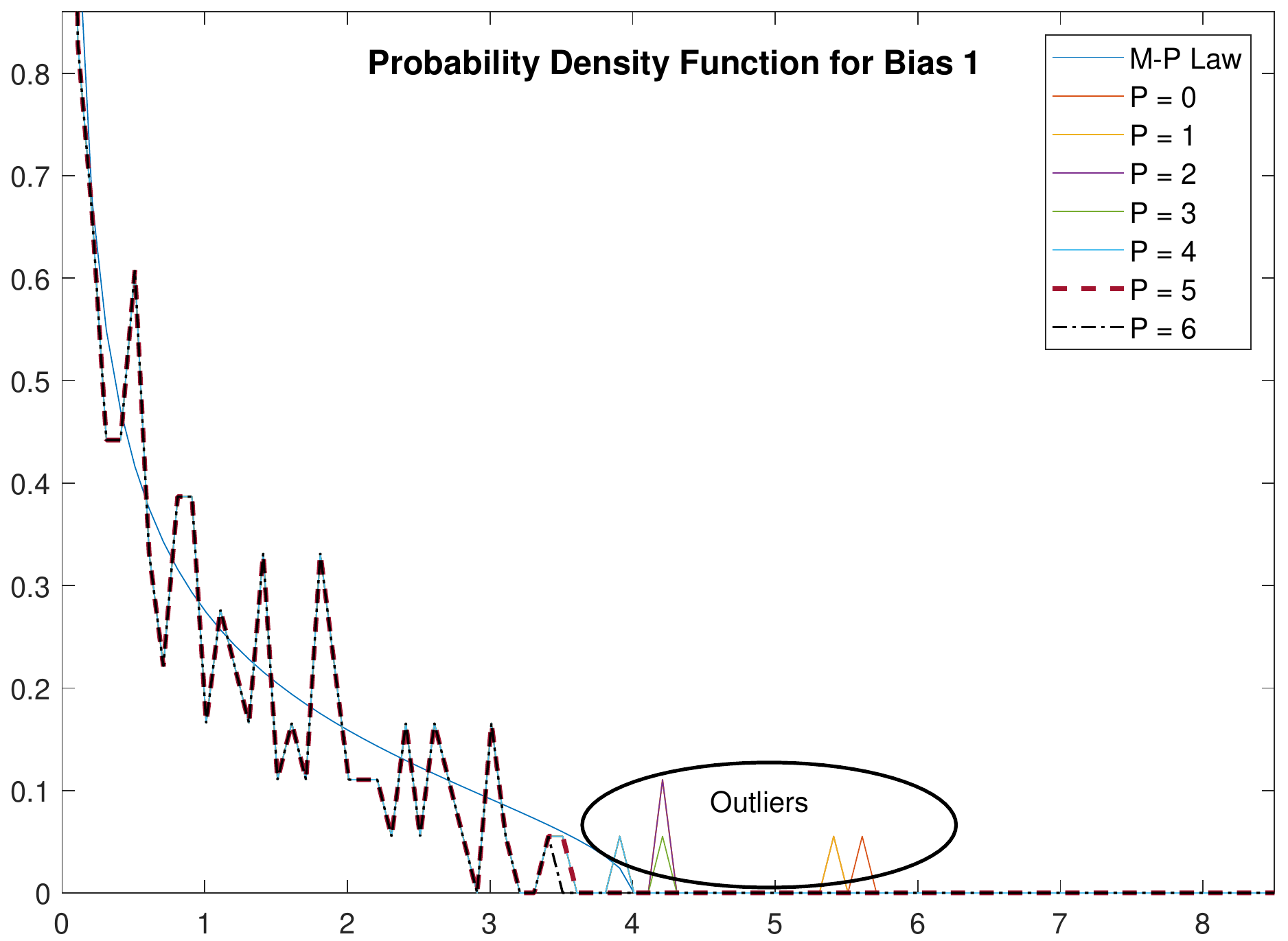}}
\subfloat[Factor Model for Bias 3]{\label{fig:fm3}
\includegraphics[width=0.24\textwidth]{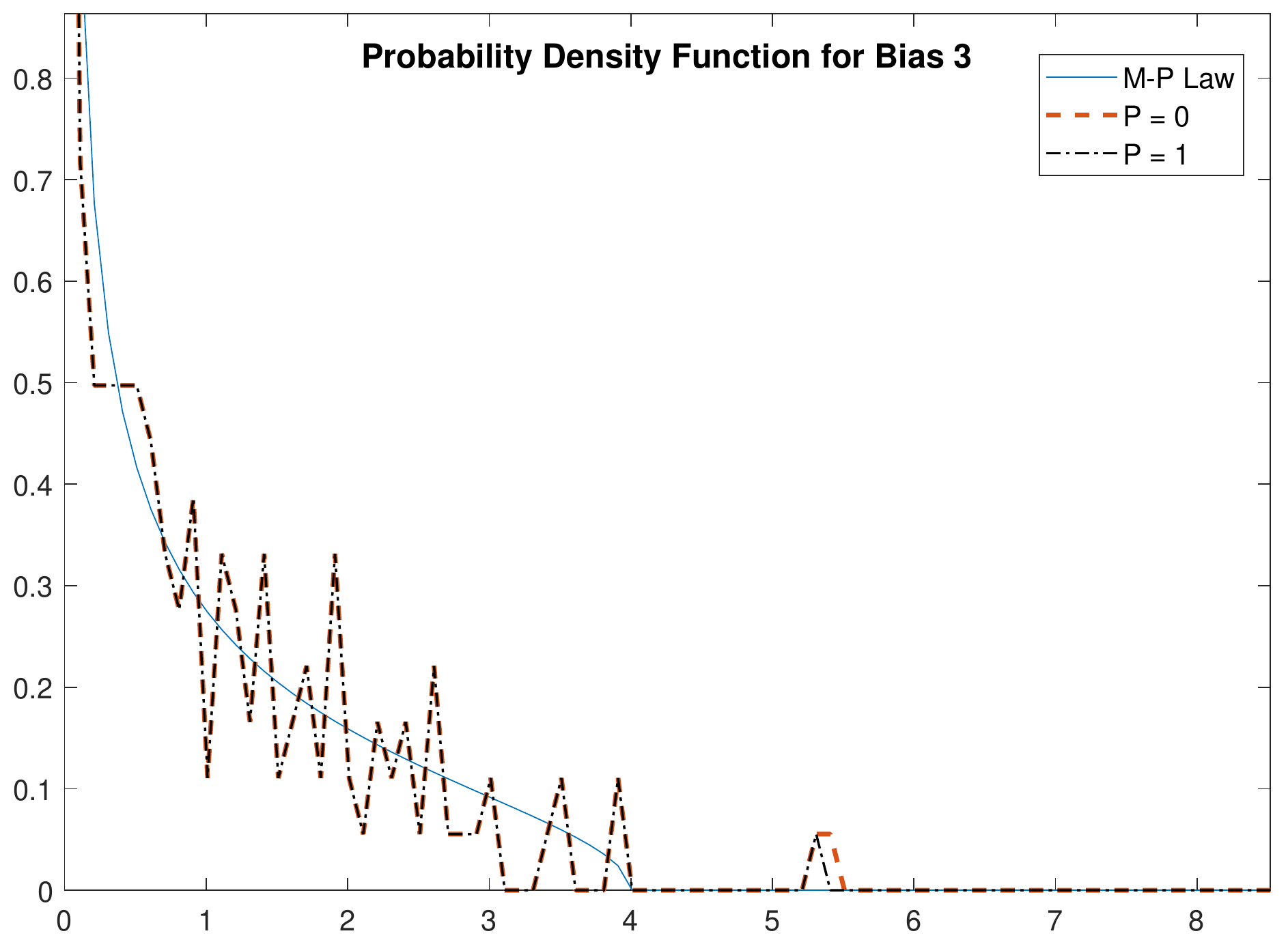}}

\caption{Estimation Result for IEEE 118-bus System}
\label{fig:case118}
\end{figure}

Data-driven  $\mathbf J$ estimation can be seen as an application of Task 2, virtual test.
Take IEEE 118-bus system as a background.
Running `case118.m' in Matpower with the raw code, we calculate the benchmark (Fig.~\ref{fig:J1180}). Via a similar process to Sec. \ref{Sec:IEEE9case}, the estimation bias is observed. In Fig.~\ref{fig:J1181}, we find some outliers exist such as Point (21, 29). According to Eq.~\eqref{eq:J1}, Point (21, 29) represents $ {\partial {{y}_{29}}}/{\partial {{x}_{21}}}$, i.e., $  {\partial {{p}_{66}}}/{\partial {{\theta}_{49}}}$.
With this clue, we check the description file (Fig.~\ref{fig:bb1}) and find that the description parameters for the branch connecting Node 49 and Node 66 are given twice, which may represents two lines exist in practice.
We make correspond modification and a better result is acquired (Fig.~\ref{fig:J1182}). Step by step, we obtain a much better result ($10^{-4}$) as Fig.~\ref{fig:J1183}.

This example reveals that a DT, \textit{with closed-loop feedbacks, is able to interact with the real physical system}.

\subsubsection{RMT-based Analytics for Residues}
{\text{\\}}

We can conduct big data analytics to our DT.
As what DT science paradigm tells in Sec. \ref{sec:BDAandAI} and \ref{sec:RMTDL}---\textit{Only in high-dimensional space do those statistical properties and benefits exist}.
Factor models, often used for dimension reduction \cite{yeo2016random}, are employed to analysis the mismatches (bias) between the model-based $\mathbf J_0$ and the DT results, based on an  intuition that \textit{the estimation bias cannot always be regarded as pure noise}---the bias does contain some statistical information especially when it is caused by poor assumptions or improper simplifications during the estimation.

From the view of factor models, the spectrum of a covariance matrix typically consists of two parts: A few spikes and the bulk. The former represents factors that mainly drive the features and the latter arises from idiosyncratic noise.
Motivated by these two parts, we consider a minimum distance between two spectrum densities---one from a covariance structure model and the other from residues.

Regarding empirical data, factor models are formulated as
\begin{equation}
\label{eq:FactM}
\mathbf{X}={{\mathbf{L}}^{\left( p \right)}}{{\mathbf{F}}^{\left( p \right)}}+\mathbf{R}.
\end{equation}
where $\mathbf{X}\in {{\mathbb{R}}^{N\times T}}$ is empirical data,    $\mathbf{F}\in {{\mathbb{R}}^{p\times T}}$ represents factors, $\mathbf{L}\in {{\mathbb{R}}^{N\times p}}$ represents factor loadings,  $p$ is the number of factors, and $\mathbf{R}\in {{\mathbb{R}}^{N\times T}}$ represents residues.

Eq.~\eqref{eq:FactM} provides us a way to decompose the real-world data into systematic information and idiosyncratic noise. Conducting RMT analysis according to \cite{8764499}, we obtain the analytics as shown in Fig. \ref{fig:fm1} and \ref{fig:fm3}. `Estimation Bias 3', which gets a much better performance than `Estimation Bias 1', has a similar statistical trend curve but much fewer outliers. And from
Fig.~\ref{fig:bb1}, we knows that  for `Estimation 1' there exist some duplications in the branch description file.
This phenomenon indicates that the estimation result is sensitive to up-to-date topology parameters.

\section{Conclusion}
This paper conduct a preliminary exploration on digital twin for power systems (PSDT).
More than a simulation tool or cyber-physical system, PSDT is characterized by seamless and active---\textit{data-driven}, \textit{real-time}, and \textit{closed-loop}---integration between digital and physical spaces. Around these attributes, we build a framework of PSDT, and then from both science paradigms and engineering practice, outlines the backgrounds, challenges, functions, technologies,  and possible directions.
Characterized by real-time data/information flow and closed-loop feedbacks, the performance of PSDT can be guaranteed, and an example is given to reveals the interaction between the DT and the real physical system via closed-loop feedbacks.

DT, mainly driven by data, has some advantages in data utilization.
DT enable us to set a quick start only with observed data, which makes DT much more accessible to engineering in practice---we no longer heavily rely on physical models.
Moreover, with the employment of RMT and Deep learning, DT is compatible with spatial-temporal data and thus can conduct big data analytics in high-dimensional space.
It means that PSDT is friendly to uncertainties and bias which are ubiquitous in a modern grid, such as Jacobian matrix estimation bias in our case studies.
Besides, DT can be evolved, in an active and self-adaptive way, with data and feedbacks accumulation.
Increasing data collection will improve the performance of DT; it is not true, however, for model-based one.

High dimension is also discussed.
Only in high-dimensional space do some statistical properties and benefits come out. For instance, we separate Jacobian matrix estimation from PF analysis; these two tasks are closely intertwined under model-based mode or in low-dimensional space.

DT is emerging and promising enabling technologies for realizing smart grids.
In this work we discuss some steady-state applications; the real-time property is much more fit for the quasi-steady-state or transient-state analysis. For instance, the NSFC's funding (2020-2022) ``Research on the Intelligent Fault Diagnosis with High Dimensional Criteria for Distribution Networks Based on the Merger of Random Matrix Theory and Deep Learning''. Beside, along the virtual test task,  there are some directions. For instance, operational strategy optimization for all parties in a VPP (virtual power plant) \cite{Qian2016Optimal}, and MAS (multi-agent system) \cite{Qian2016ResearchMAS} can be employed  to enhance the interaction of each party. PSDT can also help with the operation, dispatch, management, and electricity market in the context of power systems. In addition, our PSDT is a good reference for other industries.

\bibliographystyle{IEEEtran}
\bibliography{helx}

\normalsize{}
\end{document}